\documentclass[useAMS,usenatbib,12pt,psfig]{mnras} %,referee
\usepackage{color}
\usepackage{graphicx,amsmath}
\usepackage{rotating}
\usepackage{txfonts}
\usepackage{pdflscape,lscape}
\usepackage[T1]{fontenc}
\usepackage{aecompl}

\newcommand{\hi}{H\,{\sc i}}

\newcommand{\prin}{$^{\prime\prime}$}

\newcommand{\km}{km\,s$^{-1}$}
\newcommand{\degree}{$^{\circ}$}

\newcommand{\msolar}{M$_{\odot}$}
\newcommand{\mstar}{M$_{*}$}

\newcommand{\mhi}{M$_{\mathrm {HI}}$}

\newcommand{\msolaryr}{M$_{\odot}$\,yr$^{-1}$}

\newcommand{\af}{$A_\mathit{flux}$}

\title{Dark Matter and  HI in Ultra-Diffuse Galaxy UGC\,2162}

\author [Sengupta {\it{et al.}}]{Chandreyee Sengupta,$^{1}$\thanks{e-mail:sengupta.chandreyee@gmail.com(CS)} T. C. Scott$^{2}$, Aeree Chung$^{3}$, O. Ivy Wong$^{4}$  \\
$^{1}$  Purple Mountain Observatory, No.8 Yuanhua Road, Qixia District, Nanjing 210034, China\\
$^{2}$ Institute of Astrophysics and Space Sciences (IA), Rua das Estrelas, 4150-762 Porto, Portugal\\
$^{3}$ Department of Astronomy, Yonsei University, 50 Yonsei-ro, Seodaemun-gu, Seoul, Republic of Korea\\
$^{4}$International Centre for Radio Astronomy Research (ICRAR), University of Western Australia, 35 Stirling Highway, WA 6009, Australia}

\begin{document}

\date{Received  ; accepted  }
\date{}
\pagerange{\pageref{firstpage}--\pageref{lastpage}} \pubyear{}

\maketitle

\label{firstpage}

\begin{abstract}
\textcolor{black}{Our GMRT \hi\ observations of the ultra diffuse galaxy (UDG) UGC\,2162, \textcolor{black}{projected $\sim$ 300 kpc from the centre of the M\,77 group,} reveal it to a have an extended \textcolor{black}{\hi\ disk (R$_{HI/R_{25}}$ $\sim$ 3.3) with a moderate rotational velocity (V$_{rot} \sim$ 31 \km). This V$_{rot}$ is  in line with that of dwarf galaxies with}  similar \hi\ mass.  We estimate an M$_{dyn}$ of $\sim$ 1.14 $\times$ 10$^{9}$ \msolar\ within the galaxy's  R$_{HI}$ $\sim$ 5.2 kpc. Additionally, our estimates \textcolor{black}{of} M$_{200}$  for the galaxy from NFW models \textcolor{black}{are} in the range of 5.0 to 8.8  $\times$ 10$^{10}$ \msolar. \textcolor{black}{Comparing UGC\,2162 to samples of UDGs with \hi\ detections show it to have amongst the  smallest R$_e$ with  its   M\hi/\mstar\  being distinctly higher and  g -- i colour slightly bluer than typical values in those samples.  We also} compared \hi\ and dark matter (DM) halo properties  of UGC\,2162 with dwarf galaxies in the LITTLE THINGS sample and find its  \textcolor{black}{ DM halo mass and profile are  within the range expected for a dwarf galaxy}. While we were  unable to to determine the origin of the galaxy's present day optical form from our study, its normal \hi\ rotation velocity in relation to its \hi\ mass, \hi\ morphology,  environment  and dwarf \textcolor{black}{mass}  DM halo ruled out some of the proposed ultra diffuse galaxy formation scenarios for this galaxy. }
\end{abstract}


 \section{Introduction}
\label{intro}

Since the discovery of forty seven ultra diffuse galaxies (UDG) in the Coma cluster by \cite{vdokkum15}, there have been several further reports of many more UDGs, mostly in galaxy clusters  \citep[e.g.,][]{yagi16,wittmann17,toloba18}. UDGs have M$_*$ of 10$^7$ to 10$^8$ \msolar,  a central g -- band surface brightness ($\mu_g$) of $>$ 24 mag arcsec$^{-2}$ and  an  effective radius \footnote{The effective radius  of a galaxy is the radius at which half of the total light of the system is emitted.} ($R_{e}$) of $\sim$ 1.5  -- 5 kpc. UDGs are found in a range of  environments  from clusters and groups to in isolation, with their relative  abundance increasing  from the field to the centres of massive galaxy clusters  \citep{vaderBurg17}. This implies a significant  role for environment in \textcolor{black}{the evolution at least some of them} \citep{carleton18}.  UDGs also display a wide range of properties from spheroids with red  g -- i colours ($\sim$ 0.8) to irregulars with  blue blue g -- i colours ($\sim$ 0.3)  \citep{trujillo2017a}.   While faint, extended, low surface brightness galaxies are not a recent discovery, the Coma UDGs revealed for the first time their relative ubiquity in a dense environment. Moreover, compared to classical LSBs, the Coma  UDGs were fainter and often more extended \citep{yagi16,vaderBurg17}. 

Despite the relatively large number of reported UDGs, little is known about their properties and formation scenario(s). Currently some of the principal ideas put forward to explain  \textcolor{black}{ UDG formation}  are: (1)  galaxies with  higher than average angular momentum  and low star formation  \citep{amorisco16} (2) failed massive galaxies where either the environment or strong stellar feedback have supressed star formation (SF) \citep{vanDokkum16, DiCintio17} (3) a subset of classical LSBs or are  a class of dwarf galaxies following an evolutionary sequence from field to cluster as suggested by \cite{yozin2015,roman17,carleton18}, i.e.,  normal dwarfs spirals which have been converted to UDGs by  an  environmentally aided process/processes (e.g.,  star formation (SF) quenching, tidal heating, tidal stripping  or ram pressure).

In this paper we present resolved Giant Metrewave Radio Telescope (GMRT) \hi\ observations of UGC\,2162, a UDG with  $\mu_g$ $\sim$24.4 mag asec$^{-2}$, $R_{e}$ $\sim$ 1.7 kpc  and a g -- i colour of 0.33$\pm$0.02  \citep{trujillo2017a}. Table \ref{table_1} gives a summary of the galaxy's properties.  \cite{trujillo2017a}  showed that the stellar component of the galaxy consists of  a elongated blue star forming region within \textcolor{black}{a} more extensive  region of low surface brightness stellar emission. Using a redshift independent distance  of 12.3$\pm$1.7 Mpc  for UGC\,2162 it is claimed to be the nearest UDG \citep{trujillo2017a}, making it an excellent candidate for a resolved \hi\ study. Being located in the IAC stripe82 Legacy Survey  region, the galaxy has deep optical images. UGC\,2162 is a member of  the M\,77 group and is projected $\sim$ 294 kpc  from the \textcolor{black}{group's} centre. The nearest group member to UGC\,2162 \textcolor{black}{with comparable or greater stellar mass  is SDSS J023848.50+003114.2 which is projected 48.7 arcmin ($\sim$ 175 kpc)  away and separated in velocity by 277 \km.}

Section \ref{obs} gives details of the GMRT observations,  with  observational results in section \ref{results}.  A discussion follows in section \ref{dis} with a summary and concluding remarks in section \ref{concl}. In this paper we adopt the redshift  to UGC\,2162 of 0.00392 and distance to the galaxy of 12.3 Mpc from \cite{trujillo2017a}. We also adopt their angular scale of 1\,arcmin $\sim$ 3.6\,kpc. All $\alpha$ and $\delta$ positions referred to throughout this paper are  J2000.0.

% With an aim to study the nature of UDGs -- their gas conent, gas fraction, star- gas correlation, dark matter content and halo mass/ sizes and to understand if there is an evolutionary sequence from isolated to cluster UDGs,  we propose to carry out an HI 21 cm imaging survey of a sample of UDGs drawn from the literature

%------------------------------------------------------------------------------------------------
% Figure 1
%------------------------------------------------------------------------------------------------
\begin{figure*}
\begin{center}
\includegraphics[ angle=0,scale=.60] {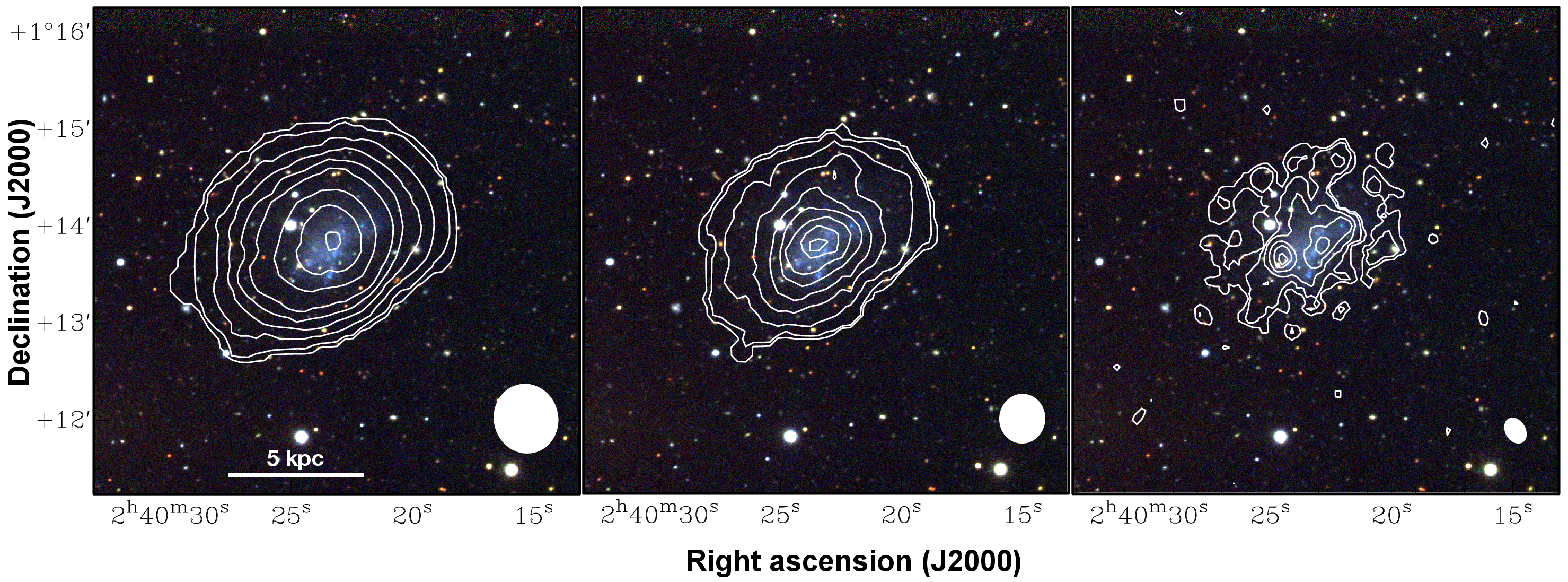}
\vspace{1cm}
\caption{\textbf{UGC\,2162:}  left to right  GMRT low (43.3\prin\ $\times$ 39.3\prin), medium (30.7\prin\ $\times$ 28.8\prin) and high (16.9\prin\ $\times$ 12.9\prin) resolution integrated \hi\  map contours on SDSS g, r, i band composite image. The GMRT beam size and orientation  is shown with white ellipses at the bottom right of each panel. {\bf Low resolution contour levels:} 10$^{20}$ atoms cm$^{-2}$ (0.3, 0.7, 1.3, 1.9, 2.6, 3.2, 4.5, 5.6, 7.1) {\bf Mid resolution contour levels:} 10$^{20}$atoms  cm$^{-2}$ (0.5, 0.9, 1.8, 3.1, 4.3, 5.6, 6.8, 8.1, 8.9) {\bf High resolution contour levels:} 10$^{20}$ atoms cm$^{-2}$ (1.2, 2.5, 5.0, 7.5, 10.0, 12.6).}
\label{fig1}
\end{center}
\end{figure*}
%------------------------------------------------------------------------------------------------

\begin{table}
\centering
\begin{minipage}{190mm}
\caption{\textcolor{black}{Properties of UGC\,2162}}
\label{table_1}
\begin{tabular}{llrr}
\hline
%Frequency & Observation  & Phase      & Phase cal    &  $\tau$    & Bandwidth &rms (per channel  & beam size   \\ 
Property\footnote{From NED}&Units&Value& \\ 
\hline
V$_{radial(optical)}$\footnote{From Hyperleda}&[\km]& 1185$\pm$6 \\
RA&[h:m:s]&02:40:23.09\\
DEC&[d:m:s]&+01:13:45.3\\
Distance\footnote{\textcolor{black}{See section 1}.}&[Mpc]&  12.3 $\pm$1.7  \\
%&& (\textcolor{black}{accepted})\\
%Spatial scale&[kpc/arcmin]&20.94 & 21.04 &NED \\
$R_{e}$\footnote{From \cite{trujillo2017a}.}& [kpc]&1.7$\pm$\textcolor{black}{0.2}\\
$\mu_g$(0)\footnote{From \cite{trujillo2017a}.}& [mag arcsec$^2$]&24.4$\pm$0.1\\
D$_{25}$ major axis  &[arcsec]&  52 \\
D$_{25}$ major axis & [kpc]& 3.2\\
Inclination\footnote{From BBarolo fit to medium resolution GMRT \hi\ cube.}& [\degree] & 55 \\
Morphology&&UDG \\
%\hi\ \aflux&&1.08 & \\
12+log(O/H)\footnote{For the bightest SF knot from a SDSS spectrum \cite{trujillo2017a}.} & &8.22$\pm$ 0.07   \\
%log($L_B$)&  [\lsolar] &14.66&14.78  \\
$M_*$\footnote{From \cite{trujillo2017a}.}   &[10$^{7}$ \msolar] &  2$\pm$2  \\
M$_{HI}$ &[10$^{7}$ \msolar]&18\\
%M$_{dyn}$ (r $<$ 4.05 kpc) &[10$^{7}$ \msolar]&90\\
%log(L$_{FIR}$) & [\lsolar]&9.53& \\
%%\hline
\end{tabular}
\end{minipage}
\end{table}

\section{Observations}
\label{obs}
UGC\,2162 was observed at 21 cm using the GMRT on the 8th March, 2018 and the observational parameters are detailed in Table \ref{table2}. The GMRT data was analysed using the standard reduction procedures with the Astronomical Image Processing System (\textsc{aips}) software package. The flux densities are on the scale of \cite{baars77}, with  uncertainties of  $\sim$ 5\%. After calibration and  continuum subtraction in the uv domain the \textsc{aips} task \textsc{imagr} was  used to convert the uv domain data to \hi\ image cubes. Finally  integrated \hi,  velocity field and velocity dispersion   maps were extracted from the image cubes using the \textsc{aips} task \textsc{momnt}. To study the  \hi\ distribution in detail,  image cubes \textcolor{black}{with} different resolutions were produced by applying different\textcolor{black}{`tapers' to the data with varying} uv limits. \textcolor{black}{Details of the final \textcolor{black}{low, medium and high}  resolution \textcolor{black}{maps}  are given in Table \ref{table2}.}

\begin{table}
\centering
\begin{minipage}{110mm}
\caption{GMRT observational and map parameters}
\label{table2}
\begin{tabular}{ll}
\hline
Rest frequency & 1420.4057 MHz \\
Observation Date &8th March, 2018\\
%Primary calibrator&?\\ 
%Phase calibrator  & \textcolor{black}{? (? mJy) } \\
%(flux density)  &   \\
Integration time  & \textcolor{black}{10.0 hrs } \\
primary beam & 24\arcmin ~at 1420.4057 MHz \\
Low resolution beam--FWHP  &43.3$^{\prime\prime}$ $\times$ 39.3$^{\prime\prime}$, PA = 13.4$^{\circ}$ \\
Medium resolution beam--FWHP& 30.7$^{\prime\prime}$ $\times$ 28.8$^{\prime\prime}$, PA = -- 4.5$^{\circ}$ \\
High resolution  beam--FWHP & 16.9$^{\prime\prime}$ $\times$ 12.9$^{\prime\prime}$, PA = 31.7$^{\circ}$ \\
%rms for low resolution map  & 1.2 mJy beam$^{-1}$  \\
%rms for medium resolution map & 0.9 mJy beam$^{-1}$  \\
%rms for high resolution map & 0.7 mJy beam$^{-1}$  \\
RA (pointing centre)&\textcolor{black}{ 02$^{\rm h}$ 40$^{\rm m}$ 23.10$^{\rm s}$  }\\
DEC (pointing centre)& \textcolor{black}{01$^\circ$ 13$^\prime$ 45.01$^{\prime\prime}$}\\
\hline
\end{tabular}
\end{minipage}
\end{table}

\section{Observational Results} 
\label{results}

\subsection{\hi\ morphology  and \textcolor{black}{\hi\ mass}} 
\label{res-morph}
Our  UGC\,2162  total intensity \hi\ maps at low, medium and high angular resolution are shown in Figure \ref{fig1}, with the  beam size and PA for each resolution  given in Table \ref{table2} and indicated with white ellipses on the maps. At the distance of 12.3 Mpc, the low, medium and high resolution  full width half power (FWHP) beams sample  the galaxy's \hi\ disc at 2.6 kpc, 1.8 kpc and 1.0 kpc respectively.  While the two lower resolution  maps show regular \hi\ morphology, the high resolution map reveals sub-structure, with the  two principal \hi\ maxima in that map having column densities   of  1.0 to 1.2 $\times$ 10$^{21}$ atoms cm$^{-2}$.   Overall the outer \hi\ disk morphology at all three resolutions  is rather symmetric and shows no signs  of extended tails or  diffuse edge structures that would signify a recent interaction. From the  low resolution \hi\ map's major axis  we estimate the R$_{HI}$ $\sim$ 86 arcsec ($\sim$5.2 kpc), compared to the R$_{25}$ = 26 arcsec (1.6 kpc) from \cite{trujillo2017a}. Its R$_{HI}$/R$_{25}$ is therefore $\sim$ 3.3, i.e. almost twice  the R$_{HI}$/R$_{25}$  of $\sim$1.8  typical of late--type galaxies \citep{broeils97}. \textcolor{black}{Comparing  the UGC\,2162 R$_{HI}$/R$_{25}$  to the morphologically unclassified Local Volume \hi\ Survey \citep[LVHIS][]{wang2017,Koribalski18} sample (limited to galaxies with M$_*$ $<$ 1 $\times$ 10$^8$ \msolar), see Figure \ref{fig2a}, indicates UGC\,2162 lies within the upper range of R$_{HI}$/R$_{25}$  for  observed galaxies  with M$_*$ $<$ 1 $\times$ 10$^8$ \msolar. }We note that the faint, patchy optical disk revealed by the deep g, r, i IAC Stripe82 composite image, Figure 1 in \cite{trujillo2017a}, extends to a radius of $\sim$ 60 arcsec (3.6 kpc), i.e. more than twice the R$_{25}$. 
%Figure 3 of \cite{trujillo2017a}  indicates \textcolor{red}{a  $\mu_g$ $>$ 26 at the outer disk edge in the IAC Stripe82  image}. 

%The relation between  the \hi\ distribution and SF is discussed  in Section \ref{dis_sf}. 

\begin{figure}
\begin{center}
\includegraphics[ angle=0,scale=.4] {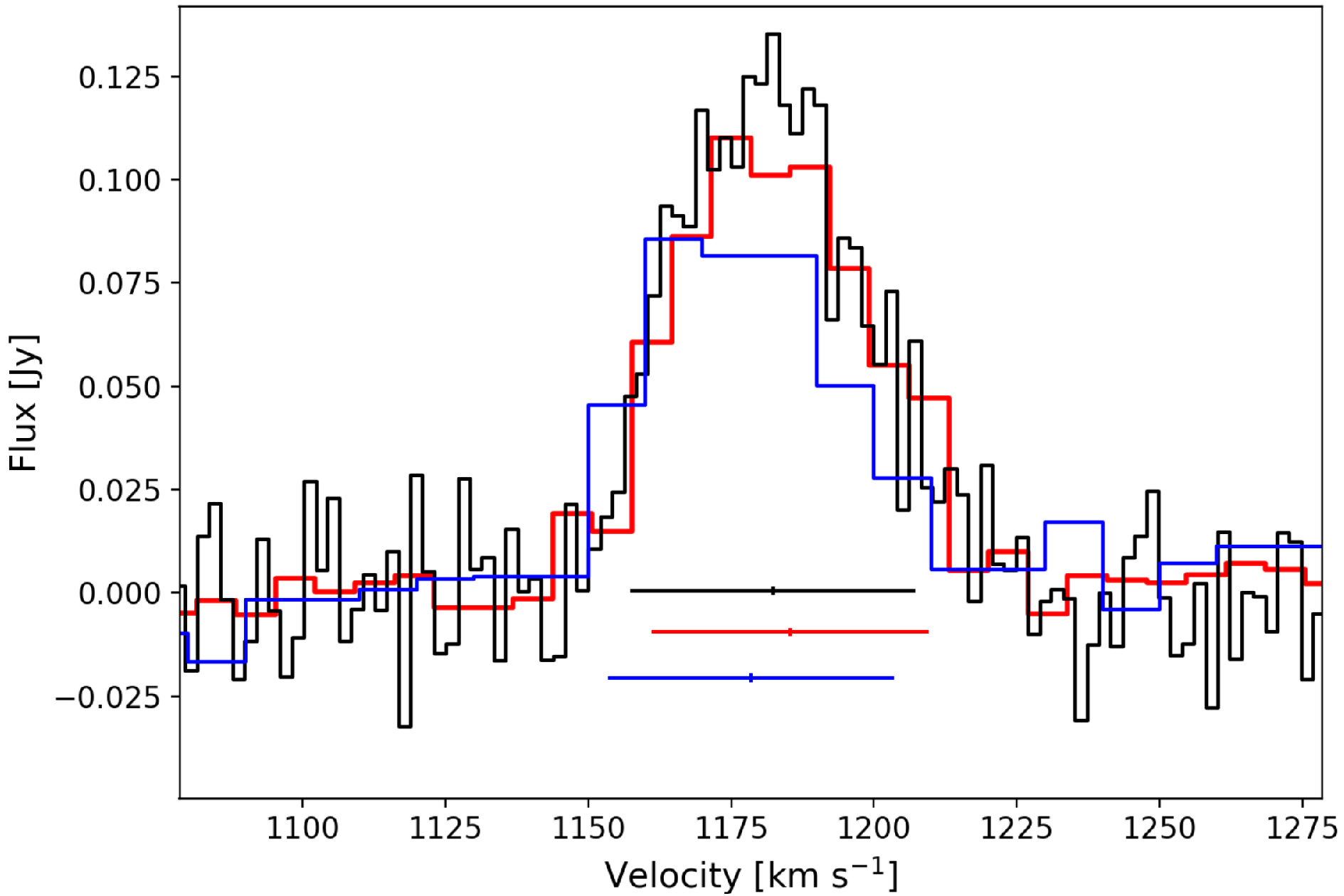}
\vspace{1cm}
\caption{\textbf{UGC\,2162  \hi\ spectra: GBT (black), GMRT (\textcolor{black}{red}) and HIPASS (blue).} The channel widths of the GBT, GMRT and HIPASS spectra are  2.1, 6.9 and 13.4  \km\ respectively The bars at the base of the spectra show the W$_{20}$ of each spectrum in its respective colour.}
\label{fig2}
\end{center}
\end{figure}

\begin{figure}
\begin{center}
\includegraphics[ angle=0,scale=.60] {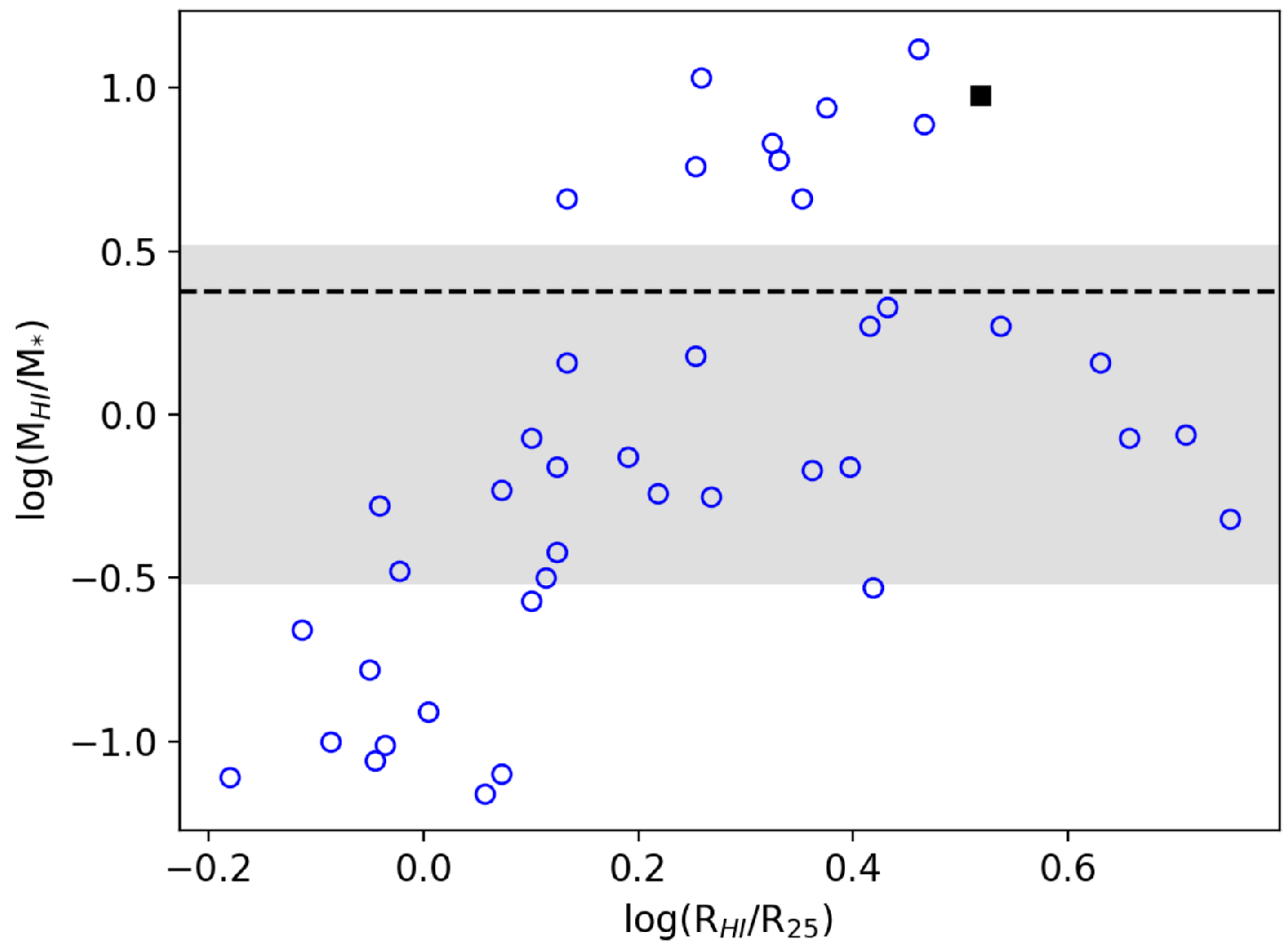}
\vspace{1cm}
\caption{\textcolor{black}{\textbf{log(M$_{HI}$/M$_*$)  v log(R$_{HI}$/R$_{25}$):}   UGC\,2162 (black square) and LVHIS local galaxies (limited to galaxies with M$_*$ $<$ 1 $\times$ 10$^8$ \msolar) from Wang et al. (2017) and Koribalski et al. (2018) - (blue circles).  The dashed line is the mean  log(M$_{HI}$/M$_*$) for 21 UDGs in the NIHAO simulation (with the shaded area showing its uncertainty) from Di Cintio et al. (2017). }}
\label{fig2a}
\end{center}
\end{figure}

In Table \ref{table3} and Figure \ref{fig2} we compare the properties of the integrated GMRT \hi\ spectrum  with those from the  42m Green Bank Telescope  \cite[GBT,][]{springob05} and \textcolor{black}{Parkes Telescope Multibeam \citep[HIPASS][]{meyer04}} single dish spectra. Both the V$_{HI}$ and W$_{20}$ derived from the three radio telescopes are in agreement  within their uncertainties. While we believe  our measurement of the W$_{20}$ from HIPASS spectrum is correct, \cite{meyer04} reported the W$_{20}$ from the HIPASS spectrum as 89 \km\ and this \textcolor{black}{is} discussed in Section \ref{dis_dyn}. The \hi\ flux  recovered by the GMRT  interferometric observation  (S$_{HI}$ = 4.5 Jy km/s) is in good agreement the GBT absorption corrected flux (S$_{HI}$ =  5.1 Jy km/s). Because it has the best signal to noise ratio (S/N)  of the available spectra, we use the GBT   flux  to estimate  the galaxy's \hi\ mass. This gives a M$_{HI}$  =  1.8 $\times$ 10$^{8}$ M$_\odot$ for the galaxy.

%  using  equation \ref{eqn1} below:\\
%\begin{equation}
%M_{HI} =2.365 \times 10^5 S_{HI} D^2\,[M_{\odot}]
%\label{eqn1}
%\end{equation}

%Where S$_{HI}$ [Jy \km]  =   \hi\ flux ($\int$ S dV), and D = distance [Mpc].  
%Applying equation  1 to the  GBT integrated flux density gives  \mhi\ = 1.8 $\times$ 10$^{8}$ M$_\odot$, which we use throughout this paper for the \hi\ mass of UGC\,2162.  

\begin{table*}
\centering
\begin{minipage}{110mm}
\caption{UGC\,2162 GMRT, GBT and HIPASS \hi\ spectra properties }
\label{table3}
\begin{tabular}{llrrrr}
\hline
Telescope&Velocity\footnote{Velocity = V$_{HI}$ except for 'Optical' which is the optical velocity from NED. }& W$_{20}$\footnote{Our measurement from the GBT and HIPASS spectra.}&Channel width& (S)/N&S$_{HI}$\\
&\km&\km&\km&&Jy km/s\\
\hline
GMRT\footnote{From the GMRT mid-resolution cube.}&1185$\pm$2&49$\pm$5&6.9&5.6&\textcolor{black}{4.5}\\
GBT&1182$\pm$1 &50$\pm$2&2.1&8.7&5.1\footnote{Self-absorption corrected, integrated flux density from \citep{springob05}.}\\
HIPASS &1178$\pm$2&53$\pm$4&13.4&6.4&5.4\footnote{From \citep{meyer04}.}\\
Optical &1185$\pm$6 && &\\
\hline
\end{tabular}
\end{minipage}
\end{table*}

\begin{figure}
\begin{center}
\includegraphics[ angle=0,scale=.60] {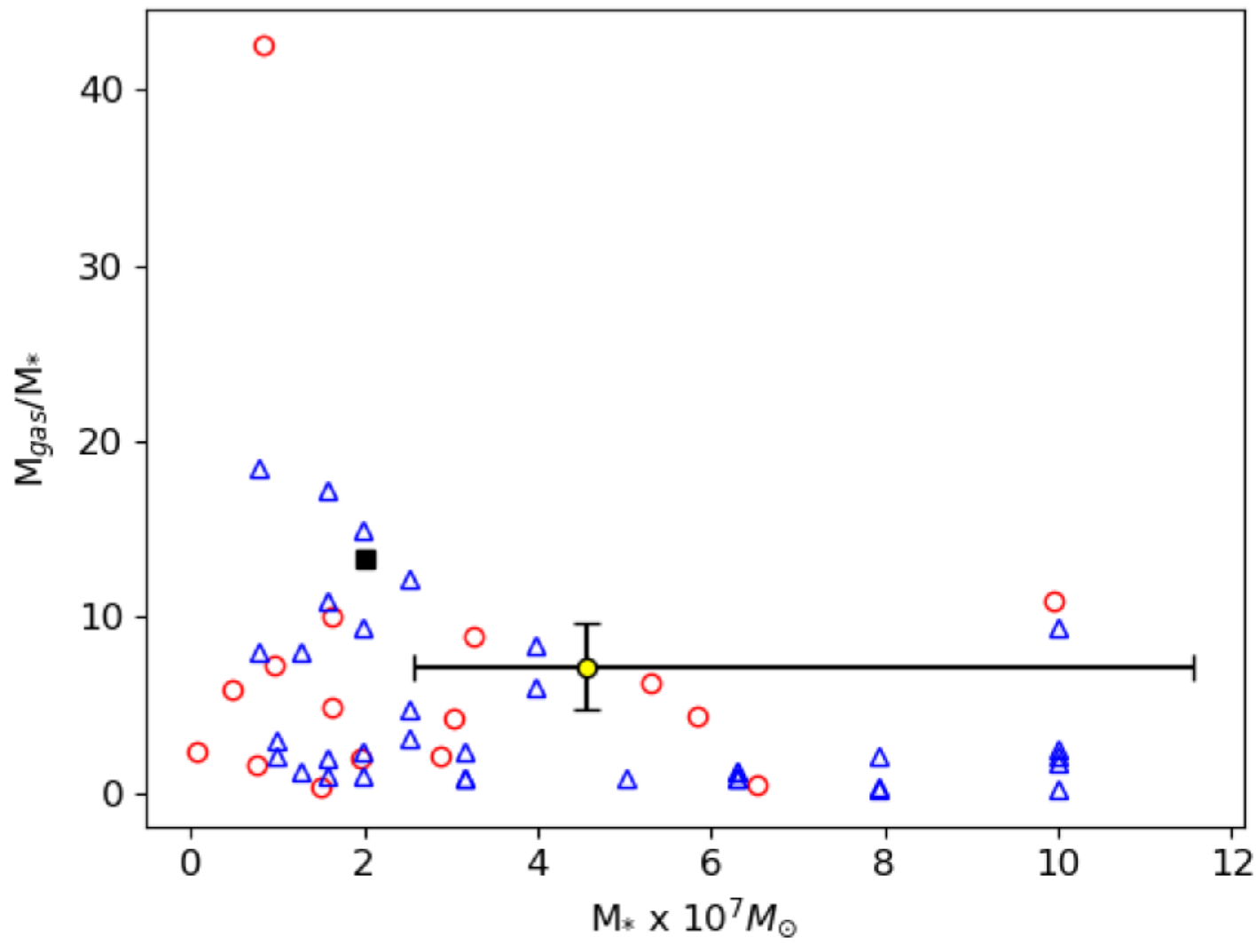}
\vspace{1cm}
\caption{\textbf{M$_{gas}$/\textcolor{black}{M$_*$  v M$_*$}}:   UGC\,2162 \textcolor{black}{(black square)}, LITTLE THINGS dwarf galaxies (red circles) from Oh et al. (2015) and LVHIS local galaxies from Wang et al. (2017) and Koribalski et al. (2018) - (blue triangles).  The yellow filled circle is the mean  M$_{gas}$/M$_*$  v M$_*$ for 21 UDGs in the NIHAO simulation  from Di Cintio et al. (2017). }
%  For UGC\,2162, and LVHIS} galaxies} M$_{gas}$ = 1.4 $\times$ \mhi\ with the additional  factor to account for molecular gas and He. }
\label{fig3}
\end{center}
\end{figure}

From the  IAC Stripe82  deep  observations   \citep{trujillo2017a} estimated the UGC\,2162 g -- band optical diameter D$_{25}$ $\sim$  52 arcsec (3.2 kpc). Using this optical diameter, the log(M$_{HI}$/${D_l}^{2}$) value for UGC\,2162 is 7.24. The average value of log(M$_{HI}$/${D_l}^{2}$) for galaxies of similar morphological type is 6.87 $\pm$ 0.17 \citep{hayn84}, indicating UGC\,2162 is \textcolor
{black}{\hi} rich compared to other galaxies of similar size and morphological type. This  conclusion is confirmed in Figure \ref{fig3} which shows M$_{gas}$/M$_*$ compared to \textcolor{black}{M$_{*}$} for UGC\,2162, the LITTLE THINGS sample of dwarf galaxies\textcolor{black}{\footnote{for which M$_*$ values available} \citep{oh15} and the morphologically unclassified LVHIS sample for galaxies, but limited to members of both samples with  M$_*$ $<$ 1 $\times$ \textcolor{black}{10$^8$ \msolar}}. \textcolor{black}{For consistency with LITTLE THINGS, M$_{gas}$ for  UGC\,2162 \textcolor{black}{and LVHIS}  = 1.4 $\times$ \mhi, with} the additional  factor to account for molecular gas \textcolor{black}{(0.04) and He (0.36)}. \textcolor{black}{ The figure shows that the M$_{gas}$/M$_*$ ratio  for UGC\,2162 is 13.3, which is almost a factor of three higher than the median M$_{gas}$/\mstar\ ratio for the LITTLE THINGS dwarfs  of 4.7 and the  mean  of 4.6$\pm$5.0 for LVHIS.} The LITTLE THINGS outlier with  M$_{gas}$/\mstar\ $>$ 40 is the  well \textcolor{black}{known} super gas rich dwarf  DDO154. \textcolor{black} Figure \ref{fig3}  also shows  the mean M$_{gas}$/M$_*$ ratio\footnote{M$_{gas}$ =M$_{HI}$ $\times$1.4 }  (7.18)  from the 21 UDGs in the NIHAO simulation \citep{DiCintio17}. From the figure we see that the UGC\,2162 M$_{gas}$/M$_*$ ratio is higher than the mean of the simulated  UDGs. (see also Figure \ref{fig2a}). Overall our analysis suggests that UGC\,2162 falls within the  ranges of  R$_{HI}$/R$_{25}$  and M$_{HI}$/M$_*$ observed  in  galaxies with similar M$_*$ in the THINGS and LVHIS samples, but  displays amongst the highest   M$_{gas}$/M$_*$ ratios in  its M$_*$ range.

\begin{figure*}
\begin{center}
\includegraphics[ angle=0,scale=.5] {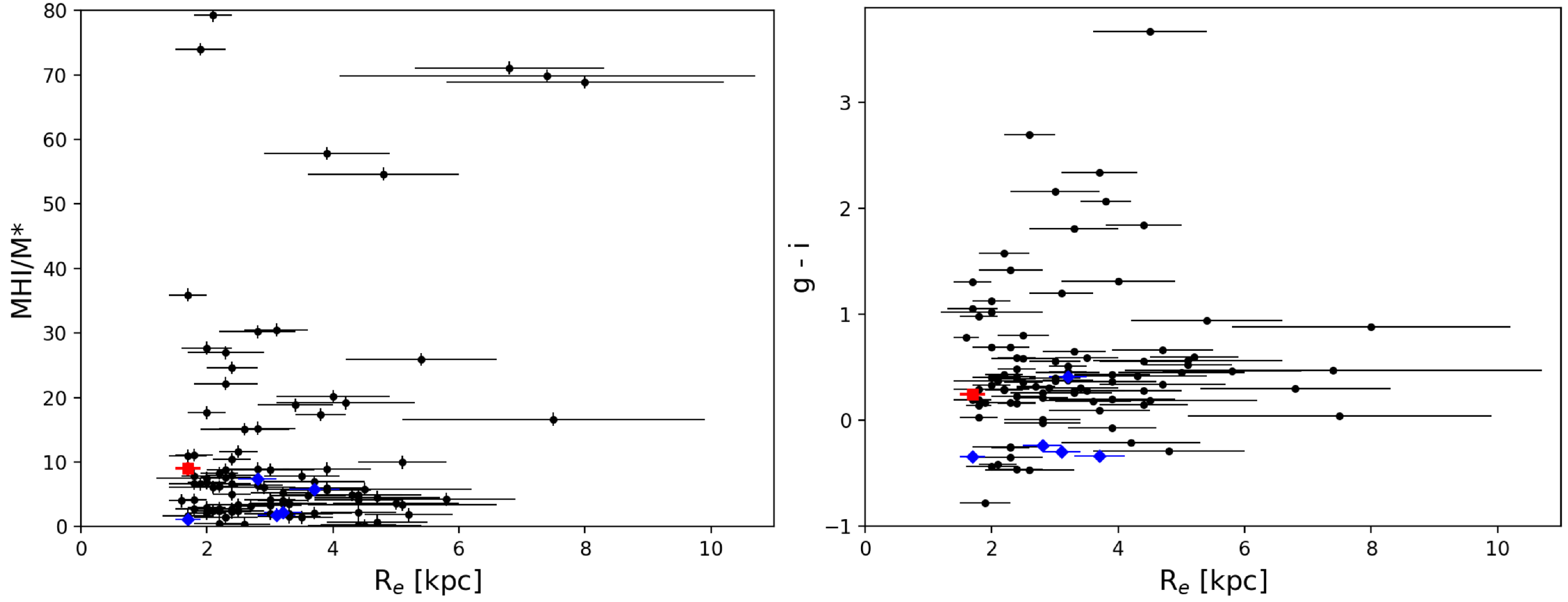}
\vspace{1cm}
\caption{\textbf{MHI/\mstar\ v R$_e$} UGC\,2162 [red square] in comparison with UDG samples with \hi\ detections;  \hi-bearing ultra-diffuse ALFALFA sources (HUDS) from Leisman et al. (2017)  [black dots] and \hi\ mapping from Spekkens and Karunakaran (2018) [blue diamonds].} 
\label{fig11}
\end{center}
\end{figure*}

\textcolor{black}{Figure \ref{fig11} compares the M\hi/\mstar\ v R$_e$ and g--i color v R$_e$ for UGC\,2162 with those from samples of UDGs with \hi\ single dish detections in ALFALFA  \citep{Leisman17} and \hi\ mapping \citep{spekkens18}. M\hi/\mstar\ for UGC\,2162 and the Spekkens sample was calculated using M$_*$  from \citep{trujillo2017a} and \cite{spekkens18} respectively.   M$_*$ was calculated for  the Leisman\footnote{The full sample contained  115 UDGs but the we excluded 19 UDGs with  M\hi/\mstar\$ $\textcolor{blue}{>}$ 80. The 19 excluded UDGs  are outliers in terms  of M\hi/\mstar\  with g band magnitudes in most cases $\geq$ 21. These galaxies were excluded because the extreme M\hi/\mstar\   \textcolor{black}{raised} questions about the reliability of their SDSS photometry for determining \mstar.} sample  using SDSS model magntudes (g and i band) and parameters  from \cite{bell03} and \cite{Blanton03}. Additionally, we calculated the SDSS \textcolor{black}{g -- i}  colours for all galaxies from their SDSS  model magnitudes, although the UGC\,2162 magnitudes were from \cite{trujillo2017a}. Magnitudes used for calculating the  g -- i colours and M$_*$ for the Leisman sample were corrected for galactic extinction using the \cite{schlafly11} values from NED extinction calculator.  From Figure \ref{fig11}  we see that UGC\,2162 has amongst the  smallest R$_e$ of the UDGs with \hi\ detections. However,  its   M\hi/\mstar\ of 9.0 while not as extreme as some Leisman UDGs is above that samples typical value.  On the other hand UGC\,2162's  g -- i colour\footnote{Corrected for galaxtic extinction.}  (0.243) is slightly bluer than most of the  Leisman sample, which probably reflects the current enhancement in SFR noted by \cite{trujillo2017a}.  The \hi\ rich nature of UGC\,2162 is also confirmed in comparisison with the simulated UDGs from \cite{DiCintio17}, see Figure \ref{fig3}. This figure shows UGC\,2162  to have a smaller R$_e$ and be gas  rich compared to an average simulated \cite{DiCintio17}  UDG. }

% \textcolor{black}{We note that  searches for molecular gas in LSB galaxies e.g. \citep{braine00} imply much lower  $\frac{H_2}{HI} $ fractions, $<$ 0.03, than are  found in normal LTGs, which is likely  linked to the lower SFRs, sSFRs and SFEs observed in LSB/UDGs. However, the upper limit  $\frac{H_2}{HI} $ fraction, 0.03, for LSBs  is only slightly  lower than the   $\frac{H_2}{HI}$   fraction of 0.04 assumed for the LITTLE THINGs dwarf galaxies in  Figures \ref{fig3} and \ref{fig3b}. So even if the UGC\,2162  $\frac{H_2}{HI}$  is substantially lower  than in the LITTLE THINGS galaxies the comparison of M$_{gas}$/\mstar\ with that sample would not be materially affected.  }

%------------------------------------------------------------------------------------------------
% Figure 2
%------------------------------------------------------------------------------------------------
%------------------------------------------------------------------------------------------------
%Thus while the stellar mass (M$_{\star}$) of the galaxy is 

%------------------------------------------------------------------------------------------------
% Figure 2a
%------------------------------------------------------------------------------------------------
%------------------------------------------------------------------------------------------------

%------------------------------------------------------------------------------------------------
% Figure 3
%------------------------------------------------------------------------------------------------
%------------------------------------------------------------------------------------------------

\nocite{oh_15}

%------------------------------------------------------------------------------------------------
% Figure 11
%------------------------------------------------------------------------------------------------
%------------------------------------------------------------------------------------------------
  \nocite{Leisman17} 
 \nocite{spekkens18}

\subsection{ \hi\ kinematics}
\label{results_kin}
\begin{figure}
\begin{center}
\includegraphics[ angle=0,scale=.65] {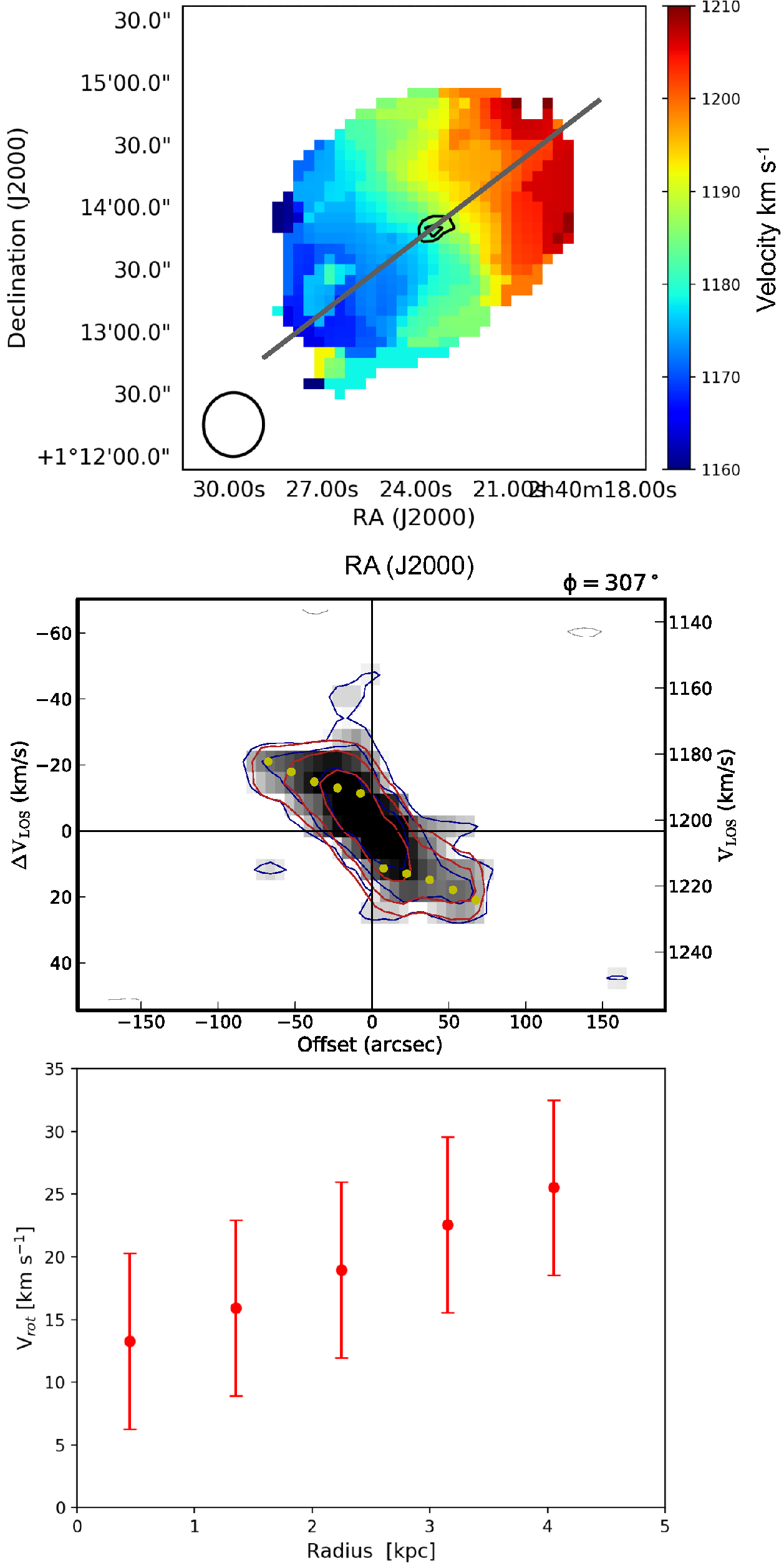}
\vspace{1cm}
\caption{\textbf{UGC\,2162  \hi\ kinematics:} \textit{Top:} GMRT velocity field from the \textcolor{black}{medium} resolution cube. The two highest column density contours from the medium resolution \hi\  integrated map are shown in black. The PV diagram (panel below) slice PA = 307\degree\ is shown with a grey line.    \textit{Middle:} PV diagram from a PA = 307 \degree slice. Positive positional offsets are to the NW. The blue contours are from the data and the red are from the \textsc{bbarolo} best fit model. \textit{Bottom:}   Rotation curve \textcolor{black}{derived from the \textsc{bbarolo} five ring  model fit.} }
\label{fig4}
\end{center}
\end{figure}

 Figure \ref{fig4} summarises the results from our  analysis of  UGC\,2162 \hi\  kinematics. The upper panel of Figure \ref{fig4} shows the \hi\ velocity field from  medium resolution cube (30.7''  $\times$ 28.9'' beam), which has the characteristics of a rotating disc \textcolor{black}{with a moderate rotation velocity and a signature of }  a warp in the SE. The black contours at the centre of this panel are the two highest \hi\ column density contours from the medium resolution total intensity \hi\ map. These \hi\ contours coincide with the kinematic centre as well as the optical centre of the galaxy. The velocity field of UGC\,2162 show no significant sign of recent major interaction.  The position velocity (PV) diagram for a slice  taken along the \hi\  major axis, position angle (PA) = 307 \degree, is shown in the middle panel of  Figure \ref{fig4} and confirms  \textcolor{black}{a}  rotating disc. A small mass of low density extra-planar \hi\ is seen  $\sim$ 15 - 20 arcsec  ($\sim$ 1 kpc) SE of the kinematic centre in the PV diagram. The lower panel of  Figure  \ref{fig4} shows the  rotation curve extracted from \textsc{bbarolo} \citep{DiTeodoro15} fit to the medium resolution cube.  The figure shows a slowly rising rotation curve  which is a common feature for dwarf galaxies. The    \textsc{bbarolo} best fit gives an  inclination of  55.5\degree, with  V$_{rot}$ at the outermost fitted ring, r = 67.5 arcsec (4.05 kpc) $\sim$ 25 \km. %\textcolor{cyan}{No change in this number due to beam correction right?}. 
\textcolor{black}{The  M$_{dyn}$  enclosed within each ring fitted with \textsc{bbarolo}  is used  in section \ref{dis_dyn} to estimate the galaxy's  virial mass (M$_{vir}$). We also estimated the V$_{rot}$ using the \hi\ W$_{20}$ (adjusted for  the inclination from the  \textsc{bbarolo} model fit) where V$_{rot}$  =  0.5 $\Delta$V [\km ]/sin(i) with $\Delta$V = W$_{20}$ (50 \km) from the GBT spectrum and i = 55.5\degree $\pi$/180. This second method  gives  V$_{rot}$  $\sim$ 30.8 \km\ at the outer edge of the detected \hi\ disk. These values are also used  in the  section \ref{dis_dyn} analysis.}%V$_{rot}$ = 25 \km\ at a radius of 67.5 arcsec , from the \textsc{bbarolo}  fit. .  

%   We also applied  3D-Based Analysis of Rotating Object via Line Observations software (\textsc{bbarolo}) from \cite{DiTeodoro15} to fit a  3D tilted-ring model to the GMRT medium resolution cube. \textsc{bbarolo} has been shown to deal well with data  with similar characteristics to our GMRT data. Althought here are relatively few beams ($\sim$5) across the medium resolution \hi\ disc  \textsc{bbarolo}'s 3D fitting  has been shown to minimise  beam smearing, which can affect 2D fitting codes such as \textsc{RORCUR}  \cite{DiTeodoro15}.  Beam smearing can cause a rotation curve  to appear shallower than it really is.

%------------------------------------------------------------------------------------------------
% Figure 4
%------------------------------------------------------------------------------------------------
%------------------------------------------------------------------------------------------------

%------------------------------------------------------------------------------------------------
% Figure 3b
%------------------------------------------------------------------------------------------------
\begin{figure}
\begin{center}
\includegraphics[ angle=0,scale=.57] {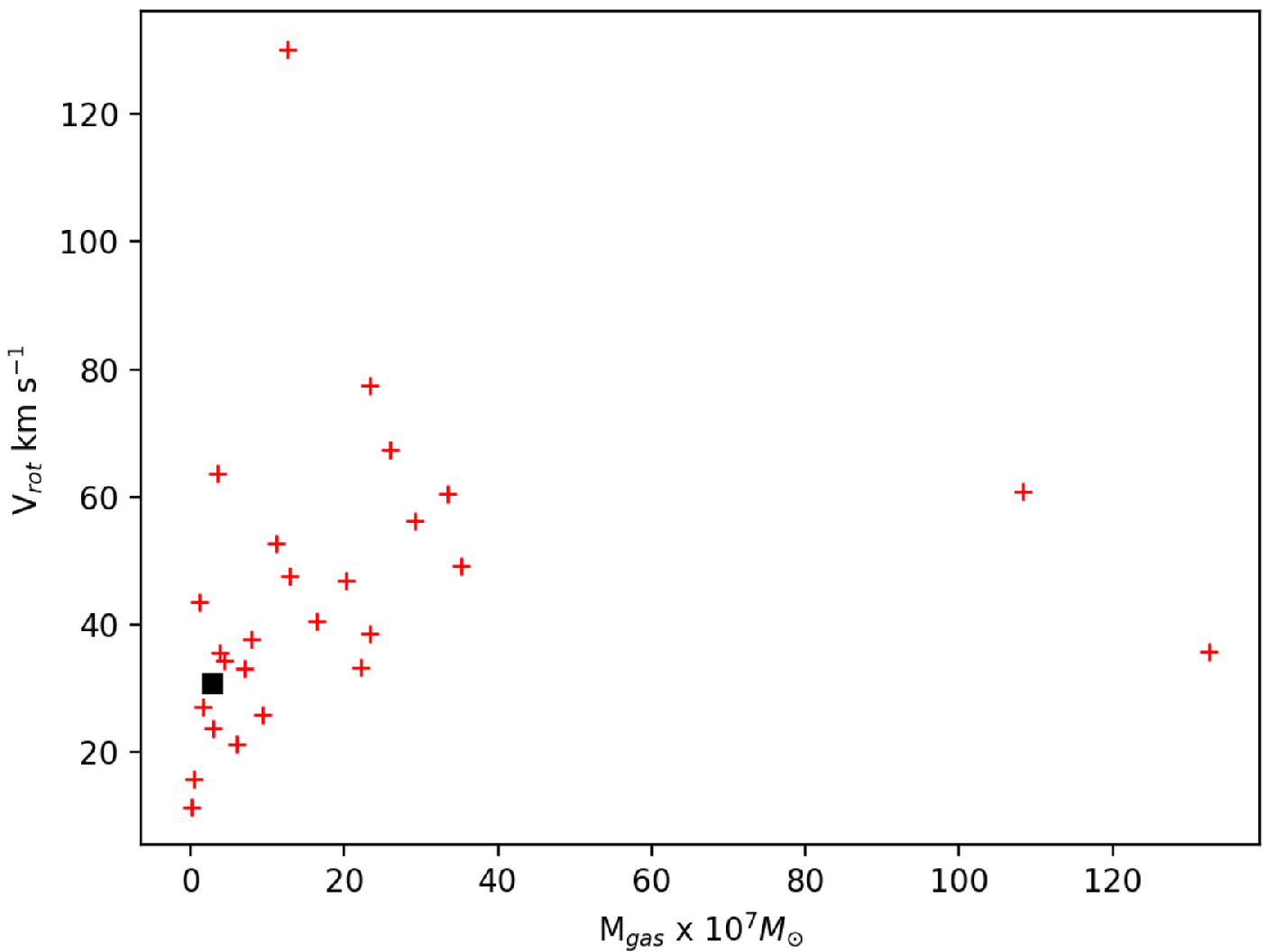}
\vspace{1cm}
\caption{\textbf{V$_{rot}$ v M$_{gas}$:  UGC\,2162 (Black square) and LITTLE THINGS dwarf galaxies (red crosses) from Oh et al. (2015)}. \textcolor{black}{The UGC\,2162, V$_{rot}$ = 30.8 \km\ is derived from the GBT W$_{20}$ in section \ref{results_kin}.   For both UGC\,2162 and the LITTLE  THINGs galaxies M$_{gas}$ =1.4 $\times$ \mhi\ with the additional  factor to account for molecular gas and He.} } 
\label{fig3b}
\end{center}
\end{figure}
%------------------------------------------------------------------------------------------------

In Figure \ref{fig3b} we compare the V$_{rot}$ from  for UGC\,2162 with the V$_{rot}$ derived from model fits to  the VLA \hi\ for  the LITTLE THINGS \textcolor{black}{dwarf} galaxies by \cite{oh15}. \textcolor{black}{The plot  shows that the  UGC\,2162 V$_{rot}$ = 30.8 \km, derived from the GBT W$_{20}$, falls within the  range of V$_{rot}$ for LITTLE THINGS  with  similar \mhi\ and below  the LITTLE THINGS median V$_{rot}$ of 39.5 \km.} 

A galaxy's   \af\ ratio is a measure of the asymmetry in its  integrated \hi\ flux density \textcolor{black}{profile}  (within its W$_{20}$ velocity range) at velocities above and below the  galaxy's systemic velocity, $V_{\mathrm {HI}}$. Even  isolated late--type galaxies  display a scatter of \af\ ratio, which is well characterised by a half Gaussian, \textcolor{black}{with its mean equal to 1.0 and a 1 $\sigma$ dispersion of 0.13. This   half  Gaussian  was obtained from a fit to the distribution of \af\ values from a sample of AMIGA\footnote{Analysis of the interstellar Medium of Isolated GAlaxies} isolated galaxies  \citep{espada11}. The value of an \af\ deviating by 1 $ \sigma$ from the mean \textcolor{black}{of that distribution} is then 1.13.} A study by  \cite{scott18} indicated that  \af\ is quite sensitive to recent ($\lesssim$ 0.7 Gyr) interactions which impact \hi\ disks. For UGC\,2162 \af\ =  1.07 $\pm$ 0.13, measured from the   GBT \hi\ spectrum, which is within the 1 $\sigma$ value from the isolated sample.  UGC\,2162's \hi\ profile is therefore  consistent with a symmetric morphology. Both the \hi\  \af\ and the velocity field indicate a relatively unperturbed rotating \hi\ disk which makes it suitable for use in determining the galaxy's M$_{dyn}$ and  M$_{vir}$. Overall UGC\,2162's  \hi\ kinematics indicate a moderately rotating disk viewed at an inclination of 55.5\degree. A comparison with the LITTLE \textcolor{black}{THINGS} sample of dwarf galaxies indicates its V$_{rot}$ of 30.8 \km\ falls within the  range of V$_{rot}$ for LITTLE THINGS  galaxies with  similar \mhi,  but below  the median V$_{rot}$ for LITTLE THINGS galaxies.

\section{Discussion}
\label{dis}

\textcolor{black}{The main goals of our \hi\ observations of UDGs \textcolor{black}{ are} to improve \textcolor{black}{knowledge of} the nature of their dark matter halos  and to \textcolor{black}{constrain UDG} formation scenarios. In this section we consider our results for UGC\,2162.}

%\subsection{Star formation and stellar mass.}
%\label{dis_sf}
%As shown in Figure \ref{fig3} and described Section \ref{res-morph}  the M$_{gas}$/\mstar\  for UGC\,2162  is a factor 3 higher than the median ratio for the LIITLE THINGs dwarf galaxies of similar M$_{gas}$.  We plot  in Figure  \ref{fig5} the bayonic and stellar mass v V$_{max}$ for UGC\,2162 compared to both Baryonic and Stellar Tully--Fisher relations from \cite{mcGaugh00,torres}. Making the assumption the M$_{mol gas}$ = M$_{HI}$ the plot shows that UGC\,2162 falls almost exactly on the  Baryonic Tully-Fisher relation (BTFR). However, for the Stellar Tully Fisher relation (STFR) mass relation the UGC\,2162 is significantly below the STFR, confirming its is deficient in stellar mass in comparison to its baryonic and DM halo mass. This and the comparision of V$_{rot}$ with the LITTLE THINGS galaxies (Figure \ref{fig3b}) implies that a mechanism, other than an abnormally high gas rotation velocity, has in the past  suppressed  SF to much lower levels than its currently elivated  star formation rate (SFR) of  0.01 \msolar\  estimated by \cite{trujillo2017a}.
%------------------------------------------------------------------------------------------------
% Figure 5
%------------------------------------------------------------------------------------------------
\begin{figure}
\begin{center}
\includegraphics[ angle=0,scale=.50] {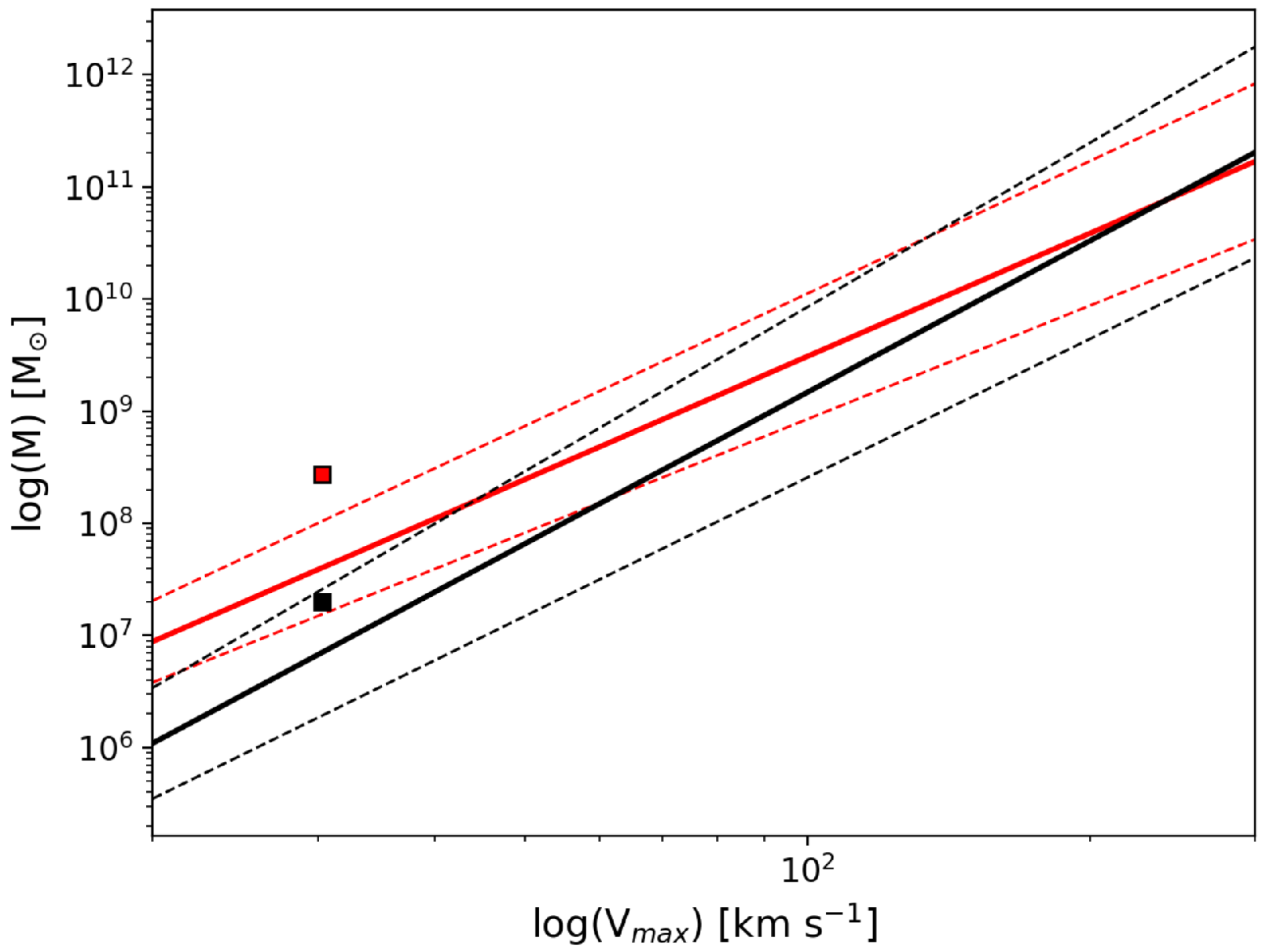}
\vspace{1cm}
\caption{\textbf{UGC\,2162 Tully--Fisher relations:}  The position of UGC\,2162 is shown with a red square  relative to the Baryonic Tully--Fisher Relation (red solid line). Also shown is UGC\,2162's position (black square) relative to the Stellar  Tully--Fisher Relation (black  solid line). The dashed lines in each colour indicate the 1 $\sigma$ scatted in their respective Tully--Fisher Relation.  }
\label{fig5}
\end{center}
\end{figure}
%------------------------------------------------------------------------------------------------

%The highest UGC\,2162 \hi\ column densities in the high resolution map approximately  coincide with the  blue star forming regions in the  SDSS background image in Figure \ref{fig1}. However, the two principal  \hi\ column density maxima in the high resolution map (x atoms cm$^{-2}$ respectively )  have counterparts  in a smoothed NUV( GALEX) image. Fig \ref{fig6} shows NUV GALEX contours on high resolution \hi. To a first order NUV matches hugest column density \hi\  What is the column density of the SF knots is it above the \citep{maybhate07} threshold? Figure \ref{fig6} show that the strongest star forming region coincides with both the \\hi column density maximum (1.26 $\times$ 10$^{21}$ atom cm$^{-2}$ in the high resolution map as well as the \hi\ velocity dispersion maximum.  

\subsection{Dark matter  halo of UGC\,2162}
\label{dis_dyn}

%The  \hi\ rotation curves of massive  spiral galaxies are observed to initially rise steeply but quickly becoming and remaining flat in the outer regions. In contrast dwarf galaxies' \hi\ rotation curves are observed to rise more slowly without necessarily flattening out \citep{DiTeodoro15}. Both types of rotation curves reflect the combined baryonic and DM mass distribution within galaxies. While, to a first order the cummulative total mass distribution in galaxies follows the Navarro--Frenk--White (NFW) profile there are important differences between the predictions of NFW models and observed rotation curves within 5 precent of the distance to the virial radius, e.g. the cusp--core problem for dwarf galaxies \citep{schaller15,oh15}. 

As a key driver of UDG  formation and evolution, their DM mass and its distribution has been debated and investigated  using models and globular cluster velocity dispersions observations \citep[e.g.,][]{vanDokkum16,toloba18} as well as  single dish \hi\  spectra \citep[e.g.][]{trujillo2017a,spekkens18}. \textcolor{black}{Seeing faint UDGs surviving in clusters, \cite{vdokkum15,vdokkum15b} predicted them to be dark matter dominated galaxies. A subsequent velocity dispersion measurement and globular cluster count for DF44 (a Coma cluster UDG, with luminosity L$_{v}$ $\sim$ 2$\times$10$^{8}$ L$_{\odot}$) indicated  a MW size \textcolor{black}{DM} halo with M$_{h}$ $\sim$ 10$^{12}$ M$_{\odot}$ \citep{vanDokkum16}.   More recent studies show that, while high dynamical to stellar mass ratios are common in UDGs, e.g. VCC1287, DF17 \textcolor{black}{and} 18 Coma UDGs, \textcolor{black}{typically} dwarf scale halos \textcolor{black}{are inferred } \cite[]{beasley16,amorisco16}. \cite{Zaritsky17}, using scaling relations, suggested that UDGs may span a range of halo masses between large spirals and dwarf galaxies, though no observations have been carried out to verify these predictions.  Additionally, several  observational studies of  UDGs have used  globular cluster velocity dispersions to estimate their DM content \citep[e.g.][]{beasley-1-16,vanDokkum18,toloba18}. \textcolor{black}{Biases} related to the use of globular \textcolor{black}{clusters} to determine UDG DM halo masses are discussed in \cite{laporte18}.  However, \textcolor{black}{resolved \hi\ can} probe DM halo properties to much larger radii than globular clusters. Therefore, at least for gas rich UDGs \hi\ is better suited to investigating the critical questions \textcolor{black}{of a UDG's DM mass and its distribution}. }

%  \hi\ is a sensitive tracer of recent  environmental or secular evolutionary processes.  \hi\ is highly sensitive to recent ($<$ 0.7 Gyr) interactions \cite{holwerda11} which means care must needs to be take when deriving DM properties from \hi\ kinematics to ensure there is no evidence of a significant recent perturbation of the \hi\ disc.

%\cite{vanDokkum16} proposed that UDGs are formed in MW mass DM halos but extreme setllar feedback suppresses their SF. Some  studies have concluded  that centres of UDGs are extremely DM dominated, e.g.   \cite{toloba18}.  Others such as \citep{beasley16,vanDokkum18} find M/L ratios consent with with massive dwarf DM halos. 

% Both types of rotation curves reflect the combined baryonic and DM mass distribution within galaxies. While, to a first order the cummulative total mass distribution in galaxies follows the Navarro--Frenk--White (NFW) profile there are important differences between the predictions of NFW models and observed rotation curves within 5 precent of the distance to the virial radius, e.g. the cusp--core problem for dwarf galaxies \citep{schaller15,oh15}.

\textcolor{black}{The  \hi\ rotation curves of massive  spiral galaxies are observed to initially rise steeply but quickly \textcolor{black}{become} flat and remain flat in the outer regions. In contrast dwarf galaxies' \hi\ rotation curves \textcolor{black}{are observed to} rise more slowly without necessarily flattening out \citep{DiTeodoro15} and this \textcolor{black}{is the }  trend observed  in the  UGC\,2162 \textcolor{black}{\hi} rotation curve, \textcolor{black}{see Figure \ref{fig4}. }} We used the V$_{rot}$ and radius  for  each of the 5 model rings  fitted by \textsc{bbarolo} to the \textcolor{black}{medium resolution} GMRT \hi\ cube together with equation \ref{eqn2} below to calculate the M$_{dyn}$ enclosed \textcolor{black}{within the radius of each  ring}:

\begin{equation}
M_{dyn} =V_{rot}^{2}  r_{HI} /G \,[M_{\odot}]
\label{eqn2}
\end{equation}

For UGC\,2162 the \textsc{bbarolo} model ring with the largest radius from the \hi\ kinematic  centre was at  r$_{HI}$ = 67.5 arcsec ($\sim$4.05 kpc).   Using \textcolor{black}{this rings's \hi\ extent  and} V$_{rot}$ = \textcolor{black}{25.5} \km, we estimate the dynamical mass (M$_{dyn}$) within the inner 4.05 kpc  using equation \ref{eqn2} as $\sim$ 0.6 $ \times$ 10$^{9}$ \msolar. Additionally, we calculated M$_{dyn}$ = \textcolor{black}{1.14 $_{- 0.47}^{+ 0.57}$}  $\times$ 10$^9$ \msolar\ enclosed within the R$_{HI}$ $\sim$  \textcolor{black}{86 \textcolor{black}{arcsec} (5.2 kpc)}, estimated from the low resolution GMRT map  and V$_{rot}$  = 30.8 \km\ derived in Section \ref{results_kin} from the GBT W$_{20}$. \textcolor{black}{Our M$_{dyn}$ value (1.14   $\times$ 10$^9$ \msolar) enclosed within the 5.2 kpc R$_{HI}$ is only  about 25\%  of the  M$_{dyn}$ = 4.6 $\times$ 10$^9$ \msolar\ based on a guessed at, but similar, R$_{HI}$ reported in \cite{trujillo2017a}. The difference is principally  due to their adoption of an inclination corrected  V$_{rot}$  = 64 \km\ based on a HIPASS W$_{20}$ = 89 \km\  from \cite{meyer04}. The Meyer value is  nearly twice the GBT W$_{20}$ of 50 $\pm$2 \km\, we have adopted. But  Table \ref{table2}  shows  our re--measurement of  the HIPASS W$_{20}$ is  in agreement, within the uncertainties, with both the GBT and GMRT  W$_{20}$ values.  }

\begin{figure}
\begin{center}
\includegraphics[ angle=0,scale=.60] {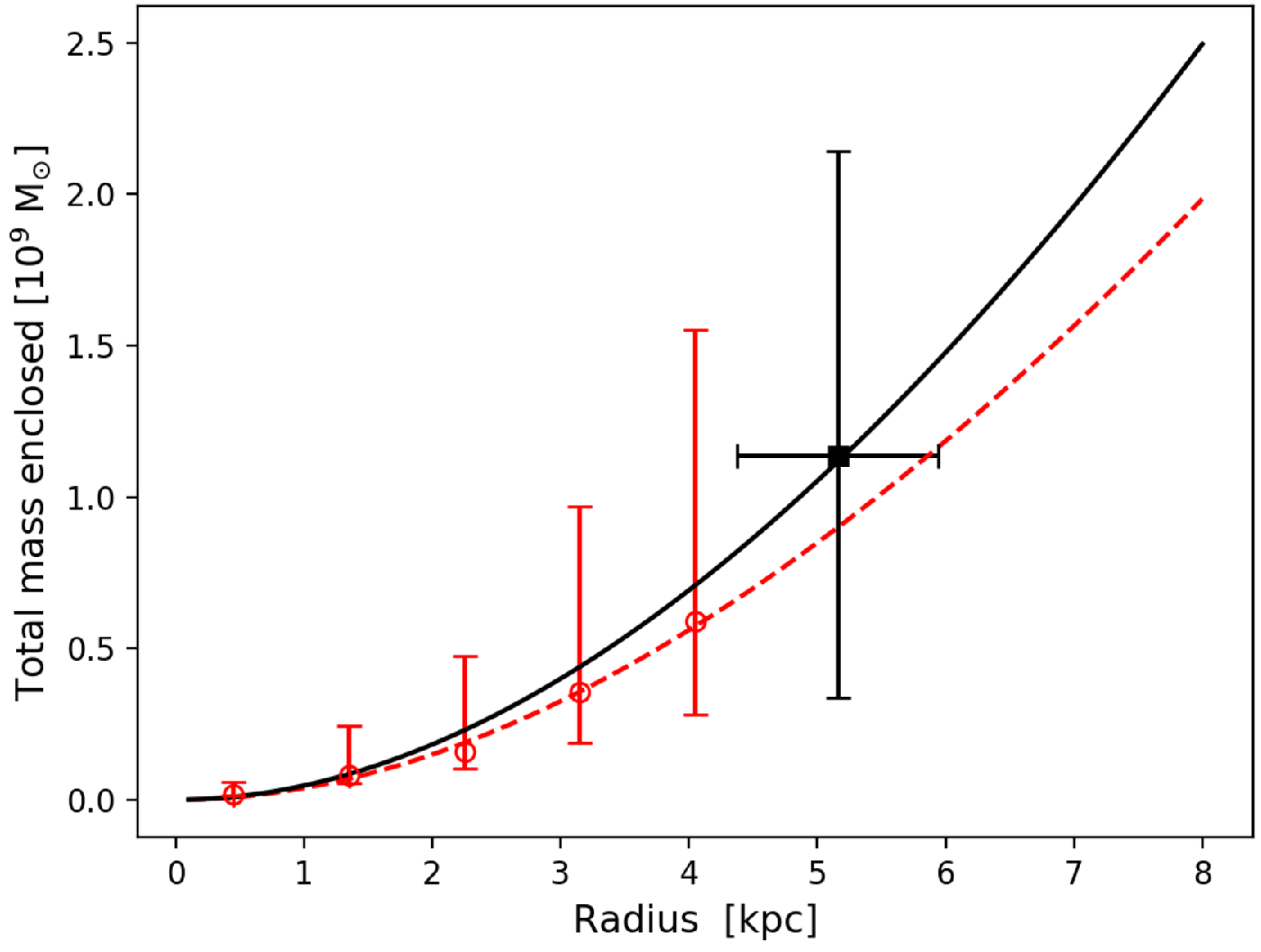}
\vspace{1cm}
\caption{\textcolor{black}{UGC\,2162: M$_{dyn}$ interior to the five ring \textsc{bbarolo} \hi\ model fits (red open circles) and M$_{dyn}$  derived  from the R$_{HI}$ from low resolution GMRT \hi\ map and W$_{20}$ (black square). Also shown are NFW model cumulative masses for DM halos  with M$_{200}$ = 5 $\times$ 10$^{10}$ \msolar\  (red dashed curve) and  M$_{200}$ = \textcolor{black}{8.8} $\times$ 10$^{10}$ \msolar\ (black solid curve). The concentration parameter = 2 for both NFW models.} }
\label{fig7}
\end{center}
\end{figure}

\begin{figure}
\begin{center}
\includegraphics[ angle=0,scale=.60] {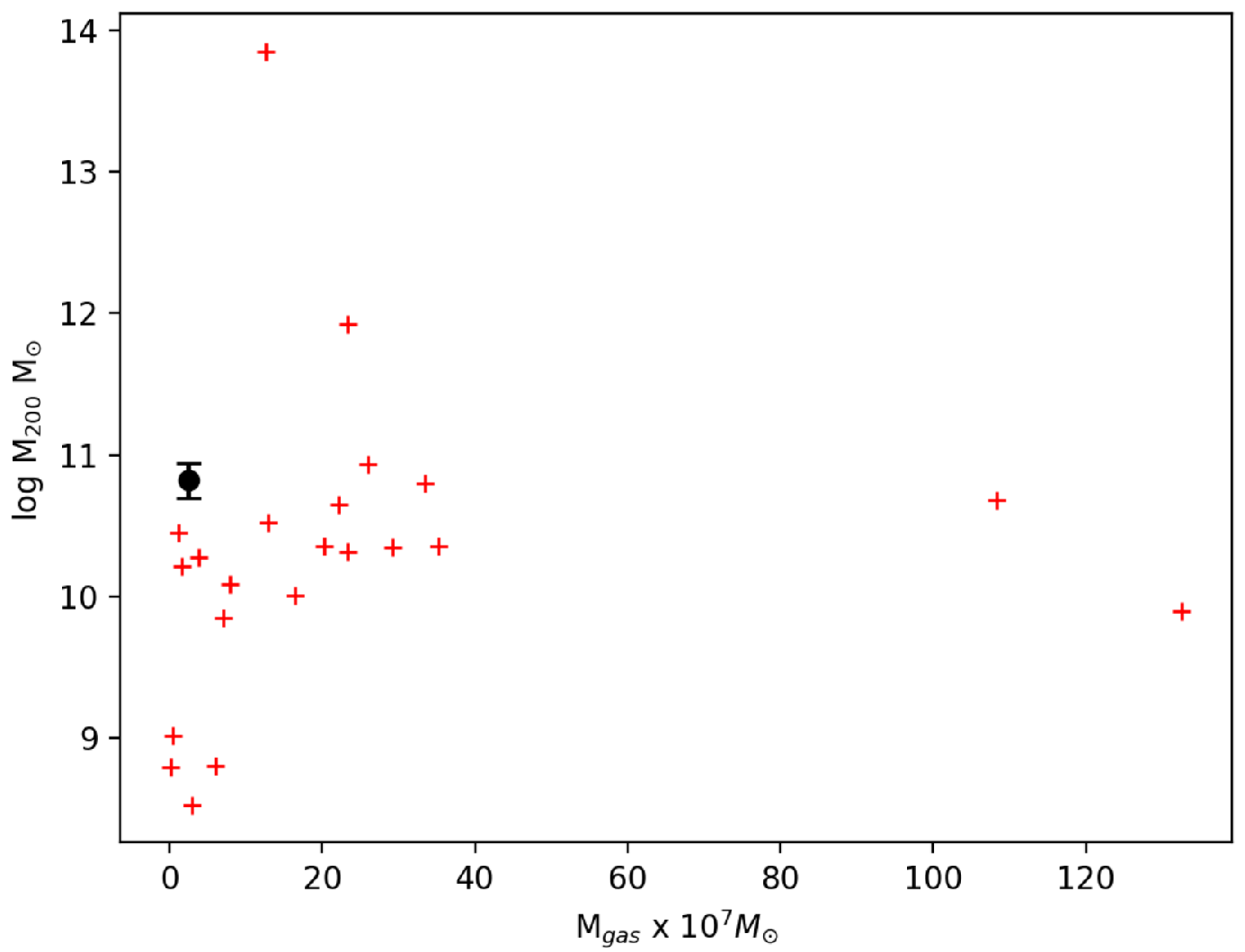}
\vspace{1cm}
\caption{ \textcolor{black}{Mean M$_{200}$ from the two NFW models  from section \ref{dis_dyn} } for UGC\,2162 (black circle)  and model fits to LITTLE THINGS dwarf galaxies from  Oh et al. (2015) (red crosses) v M$_{gas}$.
% Error bars for UGC\,2162 show the range \textcolor{red}{ $\times$ 5.0 $\times $10$^{10}$ \msolar\ to 8.8 $\times$} 10$^{10}$ \msolar. 
 }
\label{fig8}
\end{center}
\end{figure}

%This is a lower M$_{dyn}$ than estimated within the inner 5 kpc by \cite{trujillo2017a} of 4.6$\pm$0.8 $\times$ 10$^9$ \msolar. The difference is attributable to a combiation  of their larger \textcolor{black}{assumption} for the \hi\ radius and the much higher inclination uncorrected \hi\ W$_{20}$ (89\km) from \cite{meyer04}. We extracted the  HIPASS spectrum (HIPASSJ0240+01, object no. H342) refered in \cite{meyer04} at the position of UGC\,2162  in from the HIPASS on--line archive. However,  we measured a inclination uncorrected  W$_{20}$ of only  53$\pm$2 \km\ from that spectrum. Table \ref{table2} and Figure \ref{fig2} show that our measurement of the HIPASS W$_{20}$ is consistent with the HIPASS spectrum and spectra from the GMRT and GBT. 
Figure \ref{fig7} shows, with red open circles, the M$_{dyn}$ enclosed within the five \textsc{bbarolo}  model radii of 7.5", 22.5", 37.5",  52.5" and 67.5" (0.45, 1.35, 2.25, 3.15 and 4.05 kpc) from the \hi\ kinematic centre using equation \ref{eqn2}. The  M$_{dyn}$  \textcolor{black}{= 1.14 $_{- 0.47}^{+ 0.57}$ $\times$ 10$^9$ \msolar\ enclosed within the R$_{HI}$ $\sim$ (5.2 kpc)} is shown in the figure with a black square. The figure also shows \textcolor{black}{Navarro--Frenk--White (NFW)}  model\footnote{Halotools \citep{Hearin17}} cumulative masses for DM halos  with M$_{200}$ = 5 $\times$ 10$^{10}$ \msolar\  (red dashed curve) and  M$_{200}$ = \textcolor{black}{8.8} $\times$ 10$^{10}$ \msolar\ (black solid curve).  %\textcolor{cyan}{(Should we try a model for 1 point, can only red curve be enough?)}. 
The respective V$_{max}$ from these halo models are 51.1 \km\ and \textcolor{black}{64.2}  \km\ with  the \textcolor{black}{respective }  R$_{200}$s of 228 kpc and \textcolor{black}{275}  kpc. \textcolor{blue}{A} concentration parameter = 2 \textcolor{black}{was used} for both NFW models. The two NFW models fit the M$_{dyn}$ enclosed \textcolor{black}{within  ring radii} derived from \textsc{bbarolo} and  the low resolution \hi\ map disk edge \textcolor{black}{radius, provide a likely range  for the UGC\,2162 DM halo properties}.   % which have lead to the 'cusp--core controversy', i.e the   observed rotation curves   conflict with the predictions of steeply rising dwarf rotation curves from cosmological simulations \citep[e.g.][]{pontzen14,oh15,carleton18}.\textcolor{cyan}{(can we take out cusp core controversy from here?)}
Figure \ref{fig8} shows the infered   \textcolor{black}{M$_{200}$  for  UGG\,2162 (mean from two  NFW halo models) = 6.9 $\times$ 10$^{10}$ \msolar,   which is a factor of three  higher than the median M$_{200}$ (2.1 $\times$ 10$^{10}$ \msolar) for the LITTLE THINGS dwarf galaxies from model fits  carried out by \cite{oh15}. Additionally, UGC\,2162's rotation curve is consistent with the slowly rising rotation curves observed in  dwarf and LSB galaxies  \citep[e.g.][]{oh15}.}%However, as our GMRT \hi\ observation is at low resolution so we cannot rule out beam smearing as contributing  the observed shallow rotation curve, although as noted earlier the use of  \textsc{bbarolo} is expected  to  minimised the beam smearing \textcolor{blue}{\bf QUESTION: I feel the bbarolo beam smearing correction was correcting only for the spatial extent of the radius and not flux, so I feel it did not help to compensate for the shallow rotation curve, but it will cut off the rotation curve at a lower velocity. Additionally, unless we know that this flux loss due to beam smearing is significant compared to our errorbar, which may not be the case, I think we don't need to mention it}.
 
To summarise, we estimate the UGC\,2162's  M$_{dyn}$ is $\sim$ \textcolor{black}{1.14} $\times$ 10$^{9}$ \msolar\  within the R$_{HI}$  (\textcolor{black}{5.2} kpc), derived from the low resolution \hi\  map. Halo model fits to both this and M$_{dyn}$ derived from  \textsc{bbarolo} ring model fits infer  M$_{200}$ in the range  5.0 \textcolor{black}{-- 8.8  $\times$ 10$^{10}$ \msolar, which is slightly higher than the median for LITTLE THINGS galaxies.}  UGC\,2162 also displays a slowly rising  \hi\ rotation curve, typical of dwarf galaxies. UDGs encompass a range of different types of galaxies from dark matter dominated (eg. DF44: \cite{vanDokkum16} to lacking in dark matter (eg. 1052-DF2: \cite{Danieli}). In that spectrum of galaxies, a comparison of the physical  properties,  M$_{200}$, \hi\ velocity curve profiles  and  V$_{rot}$ reported for  LITTLE THINGS dwarf galaxies, lead us to conclude that the  \textcolor{black}{baryonic component of} UGC\,2162 is inhabiting a dark matter halo with a mass and profile characteristic of a dwarf galaxy.

%------------------------------------------------------------------------------------------------

 %------------------------------------------------------------------------------------------------
% Figure 8
%------------------------------------------------------------------------------------------------
%------------------------------------------------------------------------------------------------
\subsection{Formation scenario}
\label{dis_form}

\textcolor{black}{ As \textcolor{black}{noted} in Section \ref{intro}, there are multiple hypotheses in the literature regarding formation of UDGs. It is also not clear if an umbrella term UDG should be used for  these galaxies which may be in reality a collection of galaxies of all sizes, morphologies and stages of evolution with the only common factor being their extremely faint central surface brightness. We discuss here possible formation scenarios of \textcolor{black}{the} dwarf UDG, UGC\,2162.}

%From our estimates of halo mass and physical properties, we infer UDG 2162 

As shown in Figure \ref{fig3} and described in Section \ref{res-morph}  the M$_{gas}$/\mstar\  for UGC\,2162  is a factor 3 higher than the median ratio for the LIITLE THINGs dwarf galaxies of similar M$_{gas}$.  We plot  in Figure  \ref{fig5} the \textcolor{black}{UGC\,2162} bayonic and stellar \textcolor{black}{masses} vs V$_{max}$  compared to both Baryonic and Stellar Tully--Fisher relations from \cite{mcGaugh00,torres}. \textcolor{black}{For this plot we made the same assumption about the relation of M$_{gas}$ to M$_{HI}$ as we did in Section \ref{dis_dyn}, i.e. M$_{gas}$ =1.4 $\times$ M$_{HI}$.  The plot shows that while UGC\,2162's baryonic mass  falls  above the upper 1 $\sigma$ uncertaintly for Baryonic Tully-Fisher relation (BTFR), its \textcolor{black}{stellar} mass is within 1 $\sigma$ uncertaintly for the Stellar Tully Fisher relation.  No strong conclusion should be drawn based on this as the mass estimates for an individual galaxy usually has  large uncertainties. But,   this result marginally \textcolor{black}{reinforces the indications that} UGC\,2162  is gas rich relative to its stellar mass, \textcolor{black}{ suggesting} } that some mechanism has in the past  suppressed  its SF and star formation effiency \textcolor{black}{{(SFE) } to much lower levels than its currently elevated  star formation rate (SFR) of  0.01 \textcolor{black}{ \msolaryr\ }  estimated by \cite{trujillo2017a}. \textcolor{black}{\cite{wong2016} argues that the secular evolution of a combination of disk stability and disk hydrostatic pressure is the 
primary driver of  the observed \hi\ based star formation efficiency 
(SFE$_{HI}$). They find angular momentum to have only a secondary effect, concluding that for stable rotating \hi\ disks the scatter in the SFE$_{HI}$ from variation in  angular  momentum is of the order of a factor of $\sim$2.  UGC\,2612's SFE$_{HI}$ is a factor of 
4 below the average SFE$_{HI}$ of the \cite{wong2016} HIPASS--selected sample of SF galaxies. This  indicates   additional  mechanisms may \textcolor{black}{have} operated over extended periods (Gyrs) to suppress its  SFE$_{HI}$  or alternatively it recently acquired a substantial fraction of \textcolor{black}{its} current \hi\ mass.  }} %Perhaps this relates to the bit of HI 
%"beard/outflow?" that is observed in Fig 5?}}
% \textcolor{red}{Moreover, the BTFR suggests the \hi\ in UGC\,2162  is rotating in relation to other galaxies with similar M$_{baryon}$}.

Although UGC\,2162 is part of a loose group of galaxies, it is currently fairly isolated with respect to its nearest  physical  companion of similar or larger mass (SDSS J023848.50+003114.2)  projected $\sim$  175 kpc away.  The \hi\ images and kinematics of UGC\,2162 show no clear signs of recent major interaction, harassment or environment related effects that could have led to a major gas loss or significantly influence  its  morphology.  On its own the presence of UGC\,2162's \hi\ warp does not necessarily indicate a recent interaction as \hi\ warps are also seen in isolated galaxies which have been free from major tidal  interaction for several Gyrs, e.g.  \citep{2012A&A...546A..95S,portas11,scott14}, where they have been  attributed to secular processes and interactions with minor satellites. Based on the projected distance to its nearest neighbour and the relative velocity of $\sim$ 277 \km,  UGC\,2162 could have suffered a tidal interaction $\sim$ 0.3 Gyr ago with its nearest neighbour ($\sim$ less than the galaxy rotation period\footnote{T$_{rot}$ = 1.45 Gyr = 2$\pi$ R$_{HI}$/V$_{rot}$ where V$_{rot}$ =0.5$\Delta$W$_{20}$/sin(i), W$_{20}$ = 50 \km\ and i = 55\degree}), which  could  conceivably be  responsible for both the \hi\ disk warp and  currently enhanced SFR. But we argue that if such an interaction did take place,  the galaxy's current state implies that it was  most likely a minor flyby  that did not sufficiently affect UGC\, 2162's morphology to induce its transformation to an UDG.  According to the modelling by \citep{holwerda11}, HI morphology asymmetries from major merger interactions remain detectable for between   0.4 to 0.7 Gyrs. So  presumably for UGC\,2162 an encounter with a similar or larger sized or larger group member would  remain obvious in the HI morphology for at least that time. In UGC\,2162 we do not see any tidal remnants, tidal features or  any gas deficiency, which supports our argument against a recent major interaction. \textcolor{black}{Hence, any process that drove UGC\,2162 to become  a} UDG is  most likely \textcolor{black}{a long term} internal or secular one.  Figure \ref{fig3b}  compares the V$_{rot}$ of UGC\,2162 to those of LITTLE THINGS dwarfs of similar \hi\ mass and we see that UGC\,2162 is a slow rotator for its \hi\ mass, in agreement with the Tully--Fisher analysis.

 The estimated halo spin parameter ($\lambda$) using \cite{hernandez07}'s estimator for spiral galaxies gives a moderate $\lambda$ value of 0.1 for UGC\,2162. Compared to the higher spin UDGs in \cite{Leisman17} or \cite{spekkens18}, UGC\,2162 seems to qualify as a moderate to lower spin galaxy. However, we prefer not to make any claims about halo spin of UGC\,2162 as none of the assumptions made to estimate $\lambda$ in \cite{hernandez07} and \cite{spekkens18} matches UGC\,2162's properties. Instead we prefer to emphasize that most of UGC\,2162's properties strongly resemble those of normal dwarf galaxies (e.g. LITTLE THINGS sample) and more importantly the galaxy does not exhibit an abnormally extended disk compared to its DM halo mass to invoke a high spin halo formation scenario for it. From this we infer that for UGC\,2162, we can safely rule out the UDG formation scenario involving high spin halos proposed by \cite{amorisco16}.

% This seems to rule out  formation scenarios that relies on an abnormally high gas angular momentum such as as proposed by  \cite{amorisco16}. } 

\begin{figure}
\begin{center}
\includegraphics[ angle=0,scale=.35] {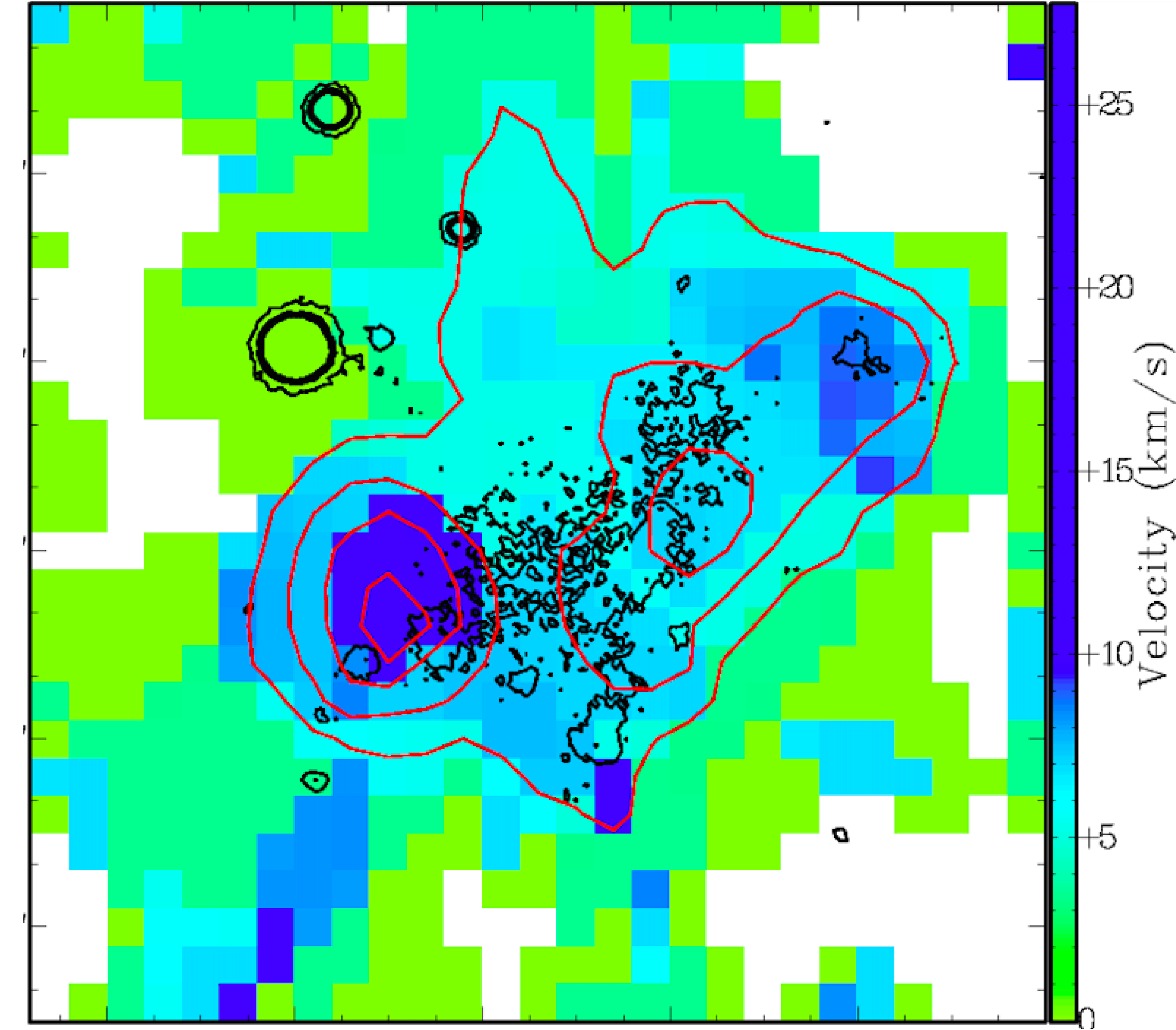}
\vspace{1cm}
\caption{\textbf{UGC\,2162 \hi\ --stellar correlation near optical center:} integrated \hi\ contours (high resolution, red) and star forming regions (SDSS g-band in black contours) plotted on \hi\ velocity dispersion (in blue/cyan) map.  }
\label{fig9}
\end{center}
\end{figure}

%This and the comparison of V$_{rot}$ with the LITTLE THINGS galaxies (Figure \ref{fig3b}) implies that 
% Additionally it is worth mentioning here that by halo mass estimates and all other physical parameters, UDG 2162 seems to be a dwarf galaxy residing in a dwarf halo and UDG 2162 thus the formation scenario of an L$_{\star}$ type failed galaxy can also be ruled out in this case. 
%UGC 2162 is a dwarf galaxy, with an  R$_{25}$ = 26 arcsec (1.6 kpc) and $R_{e}$ =28 arsec (1.7 kpc), residing in a dwarf size DM halo. 
The galaxy shows blue star forming regions near the optical centre and the highest UGC\,2162 \hi\ column densities (1.0  -- 1.3 $\times$10$^{21}$ cm$^{-2}$) in the high resolution map approximately  coincide with these  star forming zones in the  SDSS background image (Figure \ref{fig1}).  UGC\,2162 also has a  faint optical disk revealed in the IAC stripe82 \textcolor{black}{g, r, i composite} image  in Figure 1 of \cite{trujillo2017a} of radius  $\sim$ 60 arcsec  with a $\mu_g$  $>$ \textcolor{black}{26} mag arcsec$^{-2}$ at the disk edge. Fig \ref{fig9} shows the \textcolor{black}{highest} column density contours \textcolor{black}{(in red) from}  the high resolution \hi\ map  overlayed on the \textcolor{black}{high resolution}  \hi\ velocity dispersion map.  The projected positions  of the highest density \hi\  and \hi\ velocity dispersion maximum in the Figure coincide and are  offset $\sim$ 20 \textcolor{black}{arcsec (1.2 kpc)}  to the SE of the optical centre.  Addionally,  there is evidence of diffuse extra-planar \hi\ in the pv diagram (Figure \ref{fig4}) coinciding with the highest \hi\ dispersion zone. This may indicate some moderate to low level SF related outflow, but the highly disruptive past outflows proposed by  \cite{vanDokkum16} and  \cite{DiCintio17} to explain the origin of  UDGs  can be ruled out in this case\textcolor{black}{, given the galaxy is currently undergoing a heightened level of SF and its low accumulated stellar mass. } 

%Additionally, UGC 2162 appears to be more gas rich relative to its $R_{e}$  than the simulated galaxies considered in \cite{DiCintio17}
 
%The black contours \textcolor{red}{are from the  SDSS g--band image.

%  What is the column density of the SF knots is it above the \citep{maybhate07} threshold? Figure \ref{fig6} show that the strongest star forming region coincides with both the \\hi column density maximum (1.26 $\times$ 10$^{21}$ atom cm$^{-2}$ in the high resolution map as well as the \hi\ velocity dispersion maximum. 

%

%While the dwarf mass 'cored' DM halso is consistent with the  formation senario proposed by  \cite{carleton18} is is projected in the outer parts of a  group and it large gas mass are inconsistent with passage though a cluster core proposed in that senario. apply.

%------------------------------------------------------------------------------------------------
% Figure9
%------------------------------------------------------------------------------------------------
%------------------------------------------------------------------------------------------------

\section{Summary and concluding remarks}
\label{concl}
\textcolor{black}{Overall the}  \hi\ morphology and kinematics of UDG 2162 indicates a rather \textcolor{black}{symmetric} \hi\ disk within  a DM halo with a mass and profile typical of dwarf galaxies. \textcolor{black}{However, the \hi\ disk contains a warp and} a moderate,  currently elevated,  SFR of   0.01 \msolar\ \textcolor{black}{which might be attributable to an interaction with a fellow group member in last $\sim$ 0.3 Gyr}. \textcolor{black}{Comparing UGC\,2162 to samples of UDGs with \hi\ detections show it to have amongst the  smallest R$_e$ with  its   M\hi/\mstar\  being higher and  g -- i colour bluer than typical values in those samples. } \textcolor{black}{The galaxy is classified as  UDG but our investigation could not unambiguously connect its current state to any of the proposed formation scenarios  in the literature.  On the other hand we could  rule out some of the most common proposed scenarios and infer the following: 1) UGC\,2162's  DM halo is a dwarf halo typical of normal dwarf galaxies 2) the similarity of UGC\,2162's  properties to those of the LITTLE THINGS sample  does not support a scenario requiring abnormally high spin 3) the environment and  \hi\ content of the galaxy does not indicate a formation scenario depending on  a \textcolor{black}{recent} interaction or environmental assistance\textcolor{black}{; 4) we did not find evidence of recent or past highly disruptive SF driven outflows. }A detailed modelling of the galaxy's star formation history may throw more light on how this galaxy formed. }

%\textcolor{blue}{Comparing UGC\,2162 to samples of UDGs with \hi\ detections show it to have amongst the  smallest R$_e$ with  its   M\hi/\mstar\  being higher and  g -- i colour bluer than typical values in those samples. }

%\section{Summary and concluding remarks}
\label{summary}

\section{Acknowledgements} 
We thank the staff of the {\it GMRT} who have made these observations possible. The {\it GMRT} is operated by the National Centre for Radio Astrophysics of the Tata Institute of Fundamental Research.  TS  acknowledge support by Funda\c{c}\~{a}o para a Ci\^{e}ncia e a Tecnologia (FCT) through national funds (UID/FIS/04434/2013), FCT/MCTES through national funds (PIDDAC) by this grant UID/FIS/04434/2019 and by FEDER through COMPETE2020 (POCI-01-0145-FEDER-007672). TS also acknowledges the support by the fellowship SFRH/BPD/103385/2014 funded by FCT (Portugal) and POPH/FSE (EC). TS additionally  acknowledges support from DL 57/2016/CP1364/CT0009. This work was supported by FCT/MCTES through national funds (PIDDAC) by this grant PTDC/FIS-AST/29245/2017. Support for this work was provided by the National Research Foundation of Korea to the Center for Galaxy Evolution Research (No. 2010-0027910) and NRF grant No. 2018R1D1A1B07048314. This research has made use of the NASA/IPAC Extragalactic Database (NED) which is operated by the Jet Propulsion Laboratory,  California Institute of Technology, under contract with the National Aeronautics and Space Administration. This research has made use of the Sloan Digital Sky Survey (SDSS). Funding for the SDSS and SDSS-II has been provided by the Alfred P. Sloan Foundation, the Participating Institutions, the National Science Foundation, the U.S. Department of Energy, the National Aeronautics and Space Administration, the Japanese Monbukagakusho, the Max Planck Society, and the Higher Education Funding Council for England. The SDSS Web Site is http://www.sdss.org/.\textcolor{black}{This research made use of APLpy, an open-source plotting package for Python hosted at http://aplpy.github.com}. This research made use of APLpy, an open--source plotting package for Python \citep{Robitaille2012}.
The Parkes telescope is part of the Australia Telescope which is funded by the Commonwealth of Australia for operation as a National Facility managed by CSIRO.

\bibliographystyle{mnras}
\bibliography{cig}

\begin{thebibliography}{}
\makeatletter
\relax
\def\mn@urlcharsother{\let\do\@makeother \do\$\do\&\do\#\do\^\do\_\do\%\do\~}
\def\mn@doi{\begingroup\mn@urlcharsother \@ifnextchar [ {\mn@doi@}
  {\mn@doi@[]}}
\def\mn@doi@[#1]#2{\def\@tempa{#1}\ifx\@tempa\@empty \href
  {http://dx.doi.org/#2} {doi:#2}\else \href {http://dx.doi.org/#2} {#1}\fi
  \endgroup}
\def\mn@eprint#1#2{\mn@eprint@#1:#2::\@nil}
\def\mn@eprint@arXiv#1{\href {http://arxiv.org/abs/#1} {{\tt arXiv:#1}}}
\def\mn@eprint@dblp#1{\href {http://dblp.uni-trier.de/rec/bibtex/#1.xml}
  {dblp:#1}}
\def\mn@eprint@#1:#2:#3:#4\@nil{\def\@tempa {#1}\def\@tempb {#2}\def\@tempc
  {#3}\ifx \@tempc \@empty \let \@tempc \@tempb \let \@tempb \@tempa \fi \ifx
  \@tempb \@empty \def\@tempb {arXiv}\fi \@ifundefined
  {mn@eprint@\@tempb}{\@tempb:\@tempc}{\expandafter \expandafter \csname
  mn@eprint@\@tempb\endcsname \expandafter{\@tempc}}}

\bibitem[\protect\citeauthoryear{{Amorisco} \& {Loeb}}{{Amorisco} \&
  {Loeb}}{2016}]{amorisco16}
{Amorisco} N.~C.,  {Loeb} A.,  2016, \mn@doi [MNRAS] {10.1093/mnrasl/slw055},
  \href {http://cdsads.u-strasbg.fr/abs/2016MNRAS.459L..51A} {459, L51}

\bibitem[\protect\citeauthoryear{{Baars}, {Genzel}, {Pauliny-Toth}  \&
  {Witzel}}{{Baars} et~al.}{1977}]{baars77}
{Baars} J.~W.~M.,  {Genzel} R.,  {Pauliny-Toth} I.~I.~K.,   {Witzel} A.,  1977,
  A\&A, \href {http://adsabs.harvard.edu/abs/1977A%26A....61...99B} {61, 99}

\bibitem[\protect\citeauthoryear{{Beasley} \& {Trujillo}}{{Beasley} \&
  {Trujillo}}{2016}]{beasley16}
{Beasley} M.~A.,  {Trujillo} I.,  2016, \mn@doi [ApJ]
  {10.3847/0004-637X/830/1/23}, \href
  {http://cdsads.u-strasbg.fr/abs/2016ApJ...830...23B} {830, 23}

\bibitem[\protect\citeauthoryear{{Beasley}, {Romanowsky}, {Pota}, {Navarro},
  {Martinez Delgado}, {Neyer}  \& {Deich}}{{Beasley}
  et~al.}{2016}]{beasley-1-16}
{Beasley} M.~A.,  {Romanowsky} A.~J.,  {Pota} V.,  {Navarro} I.~M.,  {Martinez
  Delgado} D.,  {Neyer} F.,   {Deich} A.~L.,  2016, \mn@doi [\apj]
  {10.3847/2041-8205/819/2/L20}, \href
  {https://ui.adsabs.harvard.edu/abs/2016ApJ...819L..20B} {819, L20}

\bibitem[\protect\citeauthoryear{{Bell}, {McIntosh}, {Katz}  \&
  {Weinberg}}{{Bell} et~al.}{2003}]{bell03}
{Bell} E.~F.,  {McIntosh} D.~H.,  {Katz} N.,   {Weinberg} M.~D.,  2003, \mn@doi
  [ApJS] {10.1086/378847}, \href
  {http://adsabs.harvard.edu/abs/2003ApJS..149..289B} {149, 289}

\bibitem[\protect\citeauthoryear{{Blanton}, {Hogg}, {Bahcall}, {Brinkmann},
  {Britton}, {Connolly}, {Csabai}  \& {Fukugita}}{{Blanton}
  et~al.}{2003}]{Blanton03}
{Blanton} M.~R.,  {Hogg} D.~W.,  {Bahcall} N.~A.,  {Brinkmann} J.,  {Britton}
  M.,  {Connolly} A.~J.,  {Csabai} I.,   {Fukugita} M.,  2003, \mn@doi [ApJ]
  {10.1086/375776}, \href {http://adsabs.harvard.edu/abs/2003ApJ...592..819B}
  {592, 819}

\bibitem[\protect\citeauthoryear{{Broeils} \& {Rhee}}{{Broeils} \&
  {Rhee}}{1997}]{broeils97}
{Broeils} A.~H.,  {Rhee} M.-H.,  1997, A\&A, \href
  {http://cdsads.u-strasbg.fr/abs/1997A%26A...324..877B} {324, 877}

\bibitem[\protect\citeauthoryear{{Carleton}, {Errani}, {Cooper}, {Kaplinghat},
  {Pe{\~n}arrubia}  \& {Guo}}{{Carleton} et~al.}{2019}]{carleton18}
{Carleton} T.,  {Errani} R.,  {Cooper} M.,  {Kaplinghat} M.,  {Pe{\~n}arrubia}
  J.,   {Guo} Y.,  2019, \mn@doi [\mnras] {10.1093/mnras/stz383}, \href
  {https://ui.adsabs.harvard.edu/abs/2019MNRAS.485..382C} {485, 382}

\bibitem[\protect\citeauthoryear{{Danieli}, {van Dokkum}, {Conroy}, {Abraham}
  \& {Romanowsky}}{{Danieli} et~al.}{2019}]{Danieli}
{Danieli} S.,  {van Dokkum} P.,  {Conroy} C.,  {Abraham} R.,   {Romanowsky}
  A.~J.,  2019, \mn@doi [\apjl] {10.3847/2041-8213/ab0e8c}, \href
  {http://adsabs.harvard.edu/abs/2019ApJ...874L..12D} {874, L12}

\bibitem[\protect\citeauthoryear{{Di Cintio}, {Brook}, {Dutton}, {Macci{\`o}},
  {Obreja}  \& {Dekel}}{{Di Cintio} et~al.}{2017}]{DiCintio17}
{Di Cintio} A.,  {Brook} C.~B.,  {Dutton} A.~A.,  {Macci{\`o}} A.~V.,  {Obreja}
  A.,   {Dekel} A.,  2017, \mn@doi [\mnras] {10.1093/mnrasl/slw210}, \href
  {https://ui.adsabs.harvard.edu/abs/2017MNRAS.466L...1D} {466, L1}

\bibitem[\protect\citeauthoryear{{Di Teodoro} \& {Fraternali}}{{Di Teodoro} \&
  {Fraternali}}{2015}]{DiTeodoro15}
{Di Teodoro} E.~M.,  {Fraternali} F.,  2015, \mn@doi [MNRAS]
  {10.1093/mnras/stv1213}, \href
  {http://adsabs.harvard.edu/abs/2015MNRAS.451.3021D} {451, 3021}

\bibitem[\protect\citeauthoryear{{Espada}, {Verdes-Montenegro}, {Huchtmeier},
  {Sulentic}, {Verley}, {Leon}  \& {Sabater}}{{Espada} et~al.}{2011}]{espada11}
{Espada} D.,  {Verdes-Montenegro} L.,  {Huchtmeier} W.~K.,  {Sulentic} J.,
  {Verley} S.,  {Leon} S.,   {Sabater} J.,  2011, \mn@doi [A\&A]
  {10.1051/0004-6361/201016117}, \href
  {http://adsabs.harvard.edu/abs/2011A26A...532A.117E} {532, A117}

\bibitem[\protect\citeauthoryear{{Haynes} \& {Giovanelli}}{{Haynes} \&
  {Giovanelli}}{1984}]{hayn84}
{Haynes} M.~P.,  {Giovanelli} R.,  1984, \mn@doi [AJ] {10.1086/113573}, \href
  {http://adsabs.harvard.edu/abs/1984AJ.....89..758H} {89, 758}

\bibitem[\protect\citeauthoryear{{Hearin} et~al.,}{{Hearin}
  et~al.}{2017}]{Hearin17}
{Hearin} A.~P.,  et~al., 2017, \mn@doi [AJ] {10.3847/1538-3881/aa859f}, \href
  {http://adsabs.harvard.edu/abs/2017AJ....154..190H} {154, 190}

\bibitem[\protect\citeauthoryear{{Hernandez}, {Park}, {Cervantes-Sodi}  \&
  {Choi}}{{Hernandez} et~al.}{2007}]{hernandez07}
{Hernandez} X.,  {Park} C.,  {Cervantes-Sodi} B.,   {Choi} Y.-Y.,  2007,
  \mn@doi [MNRAS] {10.1111/j.1365-2966.2006.11274.x}, \href
  {https://ui.adsabs.harvard.edu/abs/2007MNRAS.375..163H} {375, 163}

\bibitem[\protect\citeauthoryear{{Holwerda}, {Pirzkal}, {Cox}, {de Blok},
  {Weniger}, {Bouchard}, {Blyth}  \& {van der Heyden}}{{Holwerda}
  et~al.}{2011}]{holwerda11}
{Holwerda} B.~W.,  {Pirzkal} N.,  {Cox} T.~J.,  {de Blok} W.~J.~G.,  {Weniger}
  J.,  {Bouchard} A.,  {Blyth} S.-L.,   {van der Heyden} K.~J.,  2011, \mn@doi
  [MNRAS] {10.1111/j.1365-2966.2011.18940.x}, \href
  {http://adsabs.harvard.edu/abs/2011MNRAS.416.2426H} {416, 2426}

\bibitem[\protect\citeauthoryear{{Koribalski} et~al.,}{{Koribalski}
  et~al.}{2018}]{Koribalski18}
{Koribalski} B.~S.,  et~al., 2018, \mn@doi [MNRAS] {10.1093/mnras/sty479},
  \href {http://adsabs.harvard.edu/abs/2018MNRAS.478.1611K} {478, 1611}

\bibitem[\protect\citeauthoryear{{Laporte}, {Agnello}  \& {Navarro}}{{Laporte}
  et~al.}{2019}]{laporte18}
{Laporte} C. F.~P.,  {Agnello} A.,   {Navarro} J.~F.,  2019, \mn@doi [\mnras]
  {10.1093/mnras/sty2891}, \href
  {https://ui.adsabs.harvard.edu/abs/2019MNRAS.484..245L} {484, 245}

\bibitem[\protect\citeauthoryear{{Leisman} et~al.,}{{Leisman}
  et~al.}{2017}]{Leisman17}
{Leisman} L.,  et~al., 2017, \mn@doi [ApJ] {10.3847/1538-4357/aa7575}, \href
  {http://cdsads.u-strasbg.fr/abs/2017ApJ...842..133L} {842, 133}

\bibitem[\protect\citeauthoryear{{McGaugh}, {Schombert}, {Bothun}  \& {de
  Blok}}{{McGaugh} et~al.}{2000}]{mcGaugh00}
{McGaugh} S.~S.,  {Schombert} J.~M.,  {Bothun} G.~D.,   {de Blok} W.~J.~G.,
  2000, \mn@doi [ApJL] {10.1086/312628}, \href
  {http://cdsads.u-strasbg.fr/abs/2000ApJ...533L..99M} {533, L99}

\bibitem[\protect\citeauthoryear{{Meyer} et~al.,}{{Meyer}
  et~al.}{2004}]{meyer04}
{Meyer} M.~J.,  et~al., 2004, \mn@doi [MNRAS]
  {10.1111/j.1365-2966.2004.07710.x}, \href
  {http://cdsads.u-strasbg.fr/abs/2004MNRAS.350.1195M} {350, 1195}

\bibitem[\protect\citeauthoryear{{Oh} et~al.,}{{Oh} et~al.}{2015}]{oh15}
{Oh} S.-H.,  et~al., 2015, \mn@doi [AJ] {10.1088/0004-6256/149/6/180}, \href
  {http://cdsads.u-strasbg.fr/abs/2015AJ....149..180O} {149, 180}

\bibitem[\protect\citeauthoryear{{Portas} et~al.,}{{Portas}
  et~al.}{2011}]{portas11}
{Portas} A.,  et~al., 2011, \mn@doi [ApJL] {10.1088/2041-8205/739/1/L27}, \href
  {http://cdsads.u-strasbg.fr/abs/2011ApJ...739L..27P} {739, L27}

\bibitem[\protect\citeauthoryear{{Robitaille} \& {Bressert}}{{Robitaille} \&
  {Bressert}}{2012}]{Robitaille2012}
{Robitaille} T.,  {Bressert} E.,  2012, {APLpy: Astronomical Plotting Library
  in Python}, Astrophysics Source Code Library (\mn@eprint {ascl} {1208.017})

\bibitem[\protect\citeauthoryear{{Rom{\'a}n} \& {Trujillo}}{{Rom{\'a}n} \&
  {Trujillo}}{2017}]{roman17}
{Rom{\'a}n} J.,  {Trujillo} I.,  2017, \mn@doi [MNRAS] {10.1093/mnras/stx694},
  \href {http://cdsads.u-strasbg.fr/abs/2017MNRAS.468.4039R} {468, 4039}

\bibitem[\protect\citeauthoryear{{Schlafly} \& {Finkbeiner}}{{Schlafly} \&
  {Finkbeiner}}{2011}]{schlafly11}
{Schlafly} E.~F.,  {Finkbeiner} D.~P.,  2011, \mn@doi [ApJ]
  {10.1088/0004-637X/737/2/103}, \href
  {http://adsabs.harvard.edu/abs/2011ApJ...737..103S} {737, 103}

\bibitem[\protect\citeauthoryear{{Scott} et~al.,}{{Scott}
  et~al.}{2014}]{scott14}
{Scott} T.~C.,  et~al., 2014, \mn@doi [A\&A] {10.1051/0004-6361/201423701},
  \href {http://cdsads.u-strasbg.fr/abs/2014A%26A...567A..56S} {567, A56}

\bibitem[\protect\citeauthoryear{{Scott}, {Brinks}, {Cortese}, {Boselli}  \&
  {Bravo-Alfaro}}{{Scott} et~al.}{2018}]{scott18}
{Scott} T.~C.,  {Brinks} E.,  {Cortese} L.,  {Boselli} A.,   {Bravo-Alfaro} H.,
   2018, \mn@doi [MNRAS] {10.1093/mnras/sty063}, \href
  {http://cdsads.u-strasbg.fr/abs/2018MNRAS.475.4648S} {475, 4648}

\bibitem[\protect\citeauthoryear{{Sengupta} et~al.,}{{Sengupta}
  et~al.}{2012}]{2012A&A...546A..95S}
{Sengupta} C.,  et~al., 2012, \mn@doi [\aap] {10.1051/0004-6361/201219948},
  \href {https://ui.adsabs.harvard.edu/abs/2012A&A...546A..95S} {546, A95}

\bibitem[\protect\citeauthoryear{{Spekkens} \& {Karunakaran}}{{Spekkens} \&
  {Karunakaran}}{2018}]{spekkens18}
{Spekkens} K.,  {Karunakaran} A.,  2018, \mn@doi [ApJ]
  {10.3847/1538-4357/aa94be}, \href
  {http://cdsads.u-strasbg.fr/abs/2018ApJ...855...28S} {855, 28}

\bibitem[\protect\citeauthoryear{{Springob}, {Haynes}, {Giovanelli}  \&
  {Kent}}{{Springob} et~al.}{2005}]{springob05}
{Springob} C.~M.,  {Haynes} M.~P.,  {Giovanelli} R.,   {Kent} B.~R.,  2005,
  \mn@doi [ApJS] {10.1086/431550}, \href
  {http://adsabs.harvard.edu/abs/2005ApJS..160..149S} {160, 149}

\bibitem[\protect\citeauthoryear{{Toloba} et~al.,}{{Toloba}
  et~al.}{2018}]{toloba18}
{Toloba} E.,  et~al., 2018, \mn@doi [ApJL] {10.3847/2041-8213/aab603}, \href
  {http://adsabs.harvard.edu/abs/2018ApJ...856L..31T} {856, L31}

\bibitem[\protect\citeauthoryear{{Torres-Flores}, {Epinat}, {Amram}, {Plana}
  \& {Mendes de Oliveira}}{{Torres-Flores} et~al.}{2011}]{torres}
{Torres-Flores} S.,  {Epinat} B.,  {Amram} P.,  {Plana} H.,   {Mendes de
  Oliveira} C.,  2011, \mn@doi [MNRAS] {10.1111/j.1365-2966.2011.19169.x},
  \href {http://adsabs.harvard.edu/abs/2011MNRAS.416.1936T} {416, 1936}

\bibitem[\protect\citeauthoryear{{Trujillo}, {Roman}, {Filho}  \& {S{\'a}nchez
  Almeida}}{{Trujillo} et~al.}{2017}]{trujillo2017a}
{Trujillo} I.,  {Roman} J.,  {Filho} M.,   {S{\'a}nchez Almeida} J.,  2017,
  \mn@doi [ApJ] {10.3847/1538-4357/aa5cbb}, \href
  {http://adsabs.harvard.edu/abs/2017ApJ...836..191T} {836, 191}

\bibitem[\protect\citeauthoryear{{Wang} et~al.,}{{Wang}
  et~al.}{2017}]{wang2017}
{Wang} J.,  et~al., 2017, \mn@doi [MNRAS] {10.1093/mnras/stx2073}, \href
  {http://cdsads.u-strasbg.fr/abs/2017MNRAS.472.3029W} {472, 3029}

\bibitem[\protect\citeauthoryear{{Wittmann} et~al.,}{{Wittmann}
  et~al.}{2017}]{wittmann17}
{Wittmann} C.,  et~al., 2017, \mn@doi [MNRAS] {10.1093/mnras/stx1229}, \href
  {http://cdsads.u-strasbg.fr/abs/2017MNRAS.470.1512W} {470, 1512}

\bibitem[\protect\citeauthoryear{{Wong}, {Meurer}, {Zheng}, {Heckman},
  {Thilker}  \& {Zwaan}}{{Wong} et~al.}{2016}]{wong2016}
{Wong} O.~I.,  {Meurer} G.~R.,  {Zheng} Z.,  {Heckman} T.~M.,  {Thilker} D.~A.,
    {Zwaan} M.~A.,  2016, \mn@doi [MNRAS] {10.1093/mnras/stw993}, \href
  {http://adsabs.harvard.edu/abs/2016MNRAS.460.1106W} {460, 1106}

\bibitem[\protect\citeauthoryear{{Yagi}, {Koda}, {Komiyama}  \&
  {Yamanoi}}{{Yagi} et~al.}{2016}]{yagi16}
{Yagi} M.,  {Koda} J.,  {Komiyama} Y.,   {Yamanoi} H.,  2016, \mn@doi [ApJS]
  {10.3847/0067-0049/225/1/11}, \href
  {http://cdsads.u-strasbg.fr/abs/2016ApJS..225...11Y} {225, 11}

\bibitem[\protect\citeauthoryear{{Yozin} \& {Bekki}}{{Yozin} \&
  {Bekki}}{2015}]{yozin2015}
{Yozin} C.,  {Bekki} K.,  2015, \mn@doi [MNRAS] {10.1093/mnras/stv1073}, \href
  {http://adsabs.harvard.edu/abs/2015MNRAS.452..937Y} {452, 937}

\bibitem[\protect\citeauthoryear{{Zaritsky}}{{Zaritsky}}{2017}]{Zaritsky17}
{Zaritsky} D.,  2017, \mn@doi [MNRAS] {10.1093/mnrasl/slw198}, \href
  {http://cdsads.u-strasbg.fr/abs/2017MNRAS.464L.110Z} {464, L110}

\bibitem[\protect\citeauthoryear{{van Dokkum}, {Abraham}, {Merritt}, {Zhang},
  {Geha}  \& {Conroy}}{{van Dokkum} et~al.}{2015a}]{vdokkum15}
{van Dokkum} P.~G.,  {Abraham} R.,  {Merritt} A.,  {Zhang} J.,  {Geha} M.,
  {Conroy} C.,  2015a, \mn@doi [ApJL] {10.1088/2041-8205/798/2/L45}, \href
  {http://cdsads.u-strasbg.fr/abs/2015ApJ...798L..45V} {798, L45}

\bibitem[\protect\citeauthoryear{{van Dokkum} et~al.,}{{van Dokkum}
  et~al.}{2015b}]{vdokkum15b}
{van Dokkum} P.~G.,  et~al., 2015b, \mn@doi [ApJL]
  {10.1088/2041-8205/804/1/L26}, \href
  {http://cdsads.u-strasbg.fr/abs/2015ApJ...804L..26V} {804, L26}

\bibitem[\protect\citeauthoryear{{van Dokkum} et~al.,}{{van Dokkum}
  et~al.}{2016}]{vanDokkum16}
{van Dokkum} P.,  et~al., 2016, \mn@doi [ApJL] {10.3847/2041-8205/828/1/L6},
  \href {http://cdsads.u-strasbg.fr/abs/2016ApJ...828L...6V} {828, L6}

\bibitem[\protect\citeauthoryear{{van Dokkum} et~al.,}{{van Dokkum}
  et~al.}{2018}]{vanDokkum18}
{van Dokkum} P.,  et~al., 2018, \mn@doi [ApJL] {10.3847/2041-8213/aab60b},
  \href {http://cdsads.u-strasbg.fr/abs/2018ApJ...856L..30V} {856, L30}

\bibitem[\protect\citeauthoryear{{van der Burg} et~al.,}{{van der Burg}
  et~al.}{2017}]{vaderBurg17}
{van der Burg} R.~F.~J.,  et~al., 2017, \mn@doi [A\&A]
  {10.1051/0004-6361/201731335}, \href
  {http://cdsads.u-strasbg.fr/abs/2017A%26A...607A..79V} {607, A79}

\makeatother
\end{thebibliography}

%%%%%%%%%%%%%%%%%%%%%%%%%%%%%%%%%%%%%%%%

%%%%%%%%%%%%%%%%%%%%%%%%%%%%%%%%%%%%%%%%

\end{document}